\documentclass[final,5p,times,twocolumn]{elsarticle}

\usepackage{graphicx}

\usepackage{amssymb}
 \usepackage{amsthm}

\usepackage{lineno}

\usepackage{subfigure}
\usepackage{booktabs}
\usepackage{verbatim}
\usepackage{multirow}
\usepackage{scrextend}
\usepackage[setpagesize=false]{hyperref}

\begin{document}

\begin{frontmatter}



\title{Bulk NaI(Tl) scintillation low energy events selection with the ANAIS-0 module}

\author[uz,lsc,uw]{C.~Cuesta\fnref{ca}}
\ead{ccuesta@uw.edu}
\cortext[ca]{Corresponding author}
\author[uz,lsc]{J.~ Amar\'{e}}
\author[uz,lsc]{S.~Cebri\'{a}n}
\author[uz,lsc]{E.~Garc\'{\i}a}
\author[uz,lsc]{C.~Ginestra}
\author[uz,lsc,araid]{M.~Mart\'{\i}nez}
\author[uz,lsc]{M.~A. Oliv\'{a}n}
\author[uz,lsc]{Y.~Ortigoza}
\author[uz,lsc]{A.~Ortiz~de~Sol\'{o}rzano}
\author[uz,lsc,icma]{C.~Pobes}
\author[uz,lsc]{J.~Puimed\'{o}n}
\author[uz,lsc]{M.~L.~Sarsa}
\author[uz,lsc]{J.~A.~Villar}
\author[uz,lsc]{P.~Villar}

 \address[uz]{Laboratorio de F\'{\i}sica Nuclear y Astropart\'{\i}culas, Universidad de Zaragoza, Calle Pedro Cerbuna 12, 50009 Zaragoza, Spain}
 \address[lsc]{Laboratorio Subterr\'{a}neo de Canfranc, Paseo de los Ayerbe s/n, 22880 Canfranc Estaci\'{o}n, Huesca, Spain}
\address[uw]{\emph{Present Address:} Center for Experimental Nuclear Physics and Astrophysics, and
Department of Physics, University of Washington, Seattle, WA, US}
\address[araid]{Fundaci\'{o}n ARAID, Mar\'{\i}a de Luna 11, Edificio CEEI Arag\'{o}n, 50018 Zaragoza, Spain}
\address[icma]{\emph{Present Address:} Instituto de Ciencia de Materiales de Arag\'{o}n, Universidad de Zaragoza - CSIC, Calle Pedro Cerbuna 12, 50009 Zaragoza, Spain}

\begin{abstract}
Dark matter particles scattering off target nuclei are expected to deposit very small energies in form of nuclear recoils (below 100\,keV). Because of the low scintillation efficiency for nuclear recoils as compared to electron recoils, in most of the scintillating targets considered in the search for dark matter, the region below 10\,keVee (electron equivalent energy) concentrates most of the expected dark matter signal. For this reason, very low energy threshold (at or below 2\,keVee) and very low background are required to be competitive in the search for dark matter with such detection technique. This is the case of ANAIS (Annual modulation with NaI Scintillators), which is an experiment to be carried out at the Canfranc Underground Laboratory. A good knowledge of the detector response function for real scintillation events in the active volume, a good characterization of other anomalous or noise event populations contributing in that energy range, and the development of convenient filtering procedures for the latter are mandatory in order to achieve the required low background at such a low energy. In this work we present the characteristics of different types of events observed in large size NaI(Tl) detectors, and the event-type identification techniques developed. Such techniques allow distinguishing among events associated with bulk NaI scintillation, and events related to muon interactions in the detectors or shielding, photomultiplier origin events, and analysis event fakes. We describe the specific protocols developed to build bulk scintillation events spectra from the raw data and we apply them to data obtained with one of the ANAIS prototypes, ANAIS-0. Nuclear recoil type events were also explored using data from a neutron calibration; however pulse shape cuts were found not to be effective to discriminate them from electron recoil events. The effect of the filtering procedures developed in this nuclear recoils population has been analyzed in order to properly correct cut efficiencies in dark matter analysis.

\end{abstract}

\begin{keyword}
Dark Matter \sep Annual modulation \sep Underground Physics \sep Sodium iodide scintillators


\end{keyword}

\end{frontmatter}



\section{Introduction}
\label{sec1}

The annual modulation in the detection rates could be an evidence for the presence of galactic dark matter energy depositions even in the presence of other backgrounds~\cite{freese}. The search for such an effect is of utmost interest specially in the case of using NaI(Tl) as a target because of the DAMA/LIBRA positive result: a modulation compatible with that expected for galactic halo WIMPs has been reported after 14 cycles of measurement with 9.3\,$\sigma$ statistical significance (combining the results with the previous phase of the experiment, DAMA/NaI)~\cite{DAMAphaseI,DAMAapparatus,dama}. Other experiments with gamma background rejection have obtained negative results (some of the most recent and significant negative results can be found in~\cite{luxdm,xenon100,cdms,cdms2013,supercdms14,edelweisscdms,malbek,coupp,simple}). Recently, CoGeNT experiment has reported the presence of an annual modulation in the event rate~\cite{cogent,cogent14} that could have its origin in galactic WIMPs (although different analyses result in contradictory conclusions~\cite{cogent4,davis14,KelsoIDM}), while the dark matter hints reported by CDMS-Si~\cite{cdmssi} and CRESST~\cite{cresst} experiments are more likely attributable to unaccounted for backgrounds~\cite{straussIDM}.

The difficulty in finding dark matter candidates able to explain all the present experimental results~\cite{hooper10,frandsen11,schwetz11,belli11,kelso12,frandsen12}, the dependence on the halo and WIMP models considered to compare different targets~\cite{compl,compl2,green10,green12,friedland13,nobile13},  the partial understanding of the experimental backgrounds at low energy~\cite{Kudryavtsev2010,LIBRAanswer,nygren,muons2,pardler,LIBRAanswerPardler,pardler2,LIBRAanswerPardler2,muons1,muons3,surfCresst}, and the uncertainties in the recoil energy calibration~\cite{collarY,collarCDMS,collar} which affect the interpretation of most of the available results, make highly interesting confirming the DAMA/LIBRA annual modulation observation in a model independent way. This is the goal of the ANAIS experiment, as well as other experimental efforts as DM-Ice~\cite{dmice} and KIMS~\cite{kimsnai}. The ANAIS project is intended to search for dark matter annual modulation with 250\,kg of ultrapure NaI(Tl) scintillators at the Canfranc Underground Laboratory (LSC)~\cite{ANAIStaup11,ANAISbkg,ANAISricap13} in Spain. To be successful ANAIS would require a background rate at or below 2\,cpd/keV/kg for a 2\,keVee (electron equivalent energy) threshold, and very stable working conditions. Several prototypes have been operated in order to demonstrate the detectors performance, remaining as main challenge the achievement of radiopure enough NaI(Tl) crystals~\cite{ANAIStaup13}. In this work, we present a thorough description of the event selection procedures applied to the \mbox{ANAIS-0} module, a 9.6\,kg NaI(Tl) crystal grown by Saint Gobain and similar in shape to those made for DAMA/LIBRA. In the event selection procedure based on scintillation time constants, scintillation pulses coming from a neutron calibration performed with a different NaI(Tl) crystal have been used in order to correct the calculated efficiencies obtained from gamma calibration, accounting for the slight difference in scintillation decay time~\cite{gerbier00}. Presently, two prototypes built in collaboration with Alpha Spectra Inc., Colorado, US, grown with potassium radiopure NaI powder and having 12.5\,kg mass each, are taking data at LSC for a general performance and background assessment in the ANAIS-25 setup. All the developed protocols for filtering of real bulk scintillation events in the crystals detailed in this paper are being adapted and applied to the Alpha Spectra crystals data.

In section~2 the \mbox{ANAIS-0} module and the experimental set-up will be described. In section 3 the energy calibration of \mbox{ANAIS-0} at very low energy will be presented. The trigger efficiency will be studied in section~4. The procedure followed to select the bulk NaI scintillation events at low energy will be described in section~5. Finally, the application of such a procedure to neutron calibration data will be presented in section~6.

\section{Description of ANAIS-0 module and experimental set-up at LSC}
\label{sec2}
\begin {figure}[ht]
\includegraphics[width=0.45\textwidth]{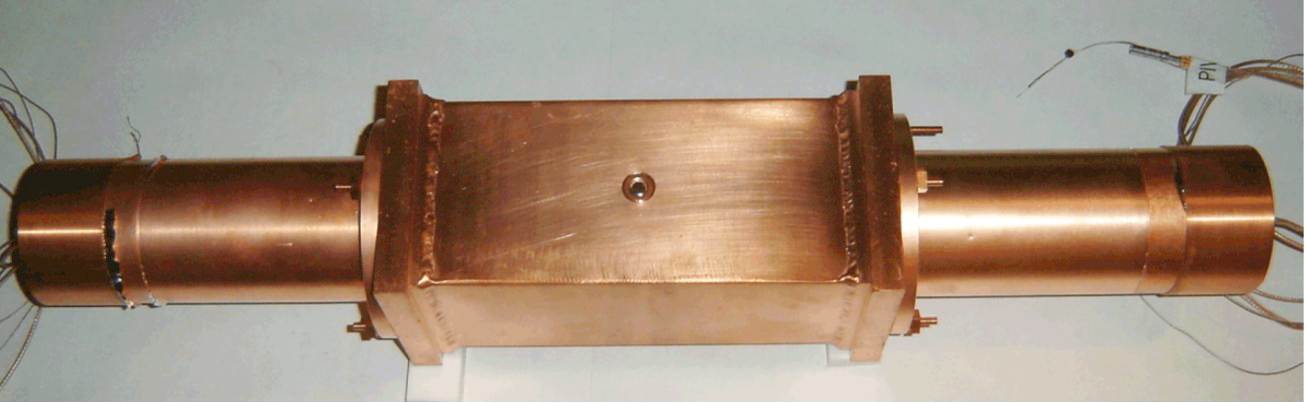}
\centering \caption{\it Picture of \mbox{ANAIS-0} module consisting of a 9.6\,kg ultrapure NaI(Tl) crystal encapsulated in ETP copper and coupled to two PMTs. The Mylar window for low-energy calibration can be observed.}
\label{fig:anais0}
\end {figure}

The detector studied throughout this work consists of a 9.6\,kg ultrapure NaI(Tl) crystal, 4"\,x\,4"\,x\,10", made by Saint-Gobain (see Figure~\ref{fig:anais0}). The crystal was encapsulated in ETP (Electrolytic Tough Pitch) copper, closing tightly the detector, and using two synthetic quartz windows to get the light out to the photomultiplier tubes (PMTs). A Teflon sheet acts as a diffuser and a Vikuiti$^{\rm TM}$ reflector layer wraps the crystal to increase light collection efficiency. This design allowed the testing of PMTs and light guides in different configurations, as described in~\cite{ANAISbkg}, although in this work we will focus on results obtained without light guides and no further reference to them will be done in the following. The crystal was encapsulated at the University of Zaragoza and the coupling of the PMTs to the quartz windows was done later at the LSC. Low background Hamamatsu PMTs of two different models, R11065SEL and R6956MOD, have been used in this work, corresponding to data sets A and B respectively (see Table~\ref{allsetupstab}). According to the manufacturer, for both PMT models the maximum spectral response should be found at 420\,nm wavelength, matching properly the emission of NaI(Tl), and being the respective spectral sensitivity ranges from 200 to 650\,nm (R11065SEL) and from 300 to 650\,nm (R6956MOD). Nominal quantum efficiencies at 420\,nm are 29\% and 33\% for the R11065SEL, and 34\% and 35\% for the R6956MOD PMT units used in this work (given by provider). Radioactivity screening of the PMTs was done at the HPGe test bench at the LSC. Results for the measured activities corresponding to the identified radioisotopes can be found in~\cite{tesisClara}, and are significantly lower (about a factor of 3) for the R11065SEL model than for the R6956MOD model. An aluminized Mylar window, 20\,$\mu$m thick and 10\,mm in diameter, in the middle of one of the lateral crystal faces allows for external calibration at very low energies (see Figure~\ref{fig:anais0}).

\begin{table}[ht]
  \caption{Available live time (LT), PMT model, and digitizer used for the data sets used throughout this work.}
\label{allsetupstab}
\begin{tabular}{llll}
\hline\noalign{\smallskip}
  Data set & PMT model & Digitizer & LT (days) \\
\noalign{\smallskip}\hline\noalign{\smallskip}
   A & R11065SEL & Tektronix scope  & 126.3\\
   B & R6956MOD & MATACQ card & \hphantom{0}31.9\\
\noalign{\smallskip}\hline
\end{tabular}
\end{table}

\begin {figure}[ht]
\includegraphics[width=0.45\textwidth]{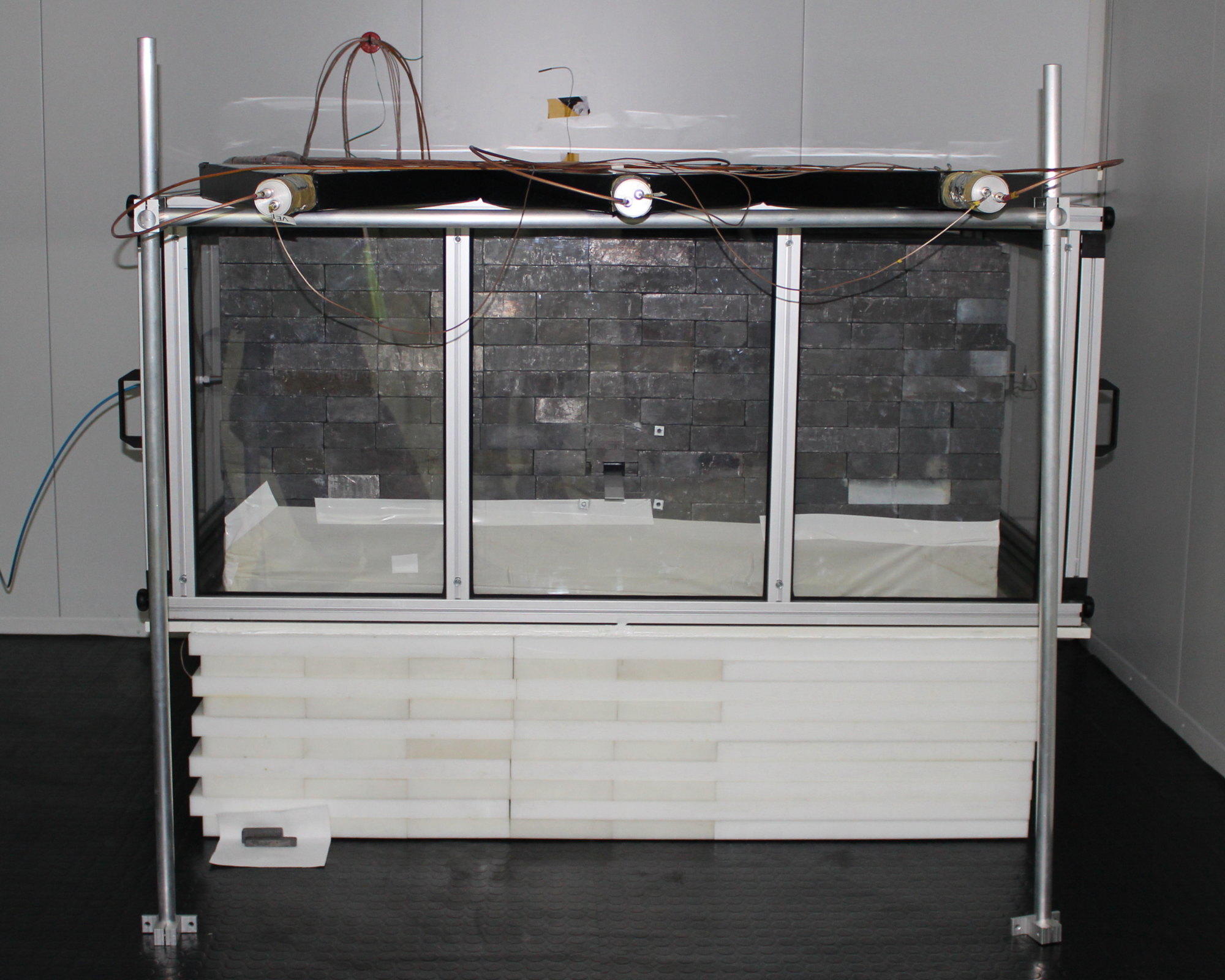}
\centering \caption{\it  \mbox{ANAIS-0} experimental layout at the new LSC facilities. The shielding consists of 10\,cm archaeological lead plus 20\,cm low activity lead, all enclosed in a PVC box tightly closed and continuously flushed with boil-off nitrogen. Active vetoes are mounted on top of the shielding to reject coincident events in \mbox{ANAIS-0}.}
\label{fig:shielding}
\end {figure}

 \mbox{ANAIS-0} module was designed to characterize and understand ANAIS background at low energy, optimize NaI scintillation events selection, fix the calibration method, and test the electronics while new more radiopure detectors are in preparation for a 250\,kg detection mass experiment. \mbox{ANAIS-0} was operated at the new LSC facilities, under a rock overburden of 2450~m.w.e. Underground operation at such a depth guarantees a significant cosmic ray suppression; measured muon flux at LSC is of the order of 10$^{-7}$\,cm$^{-2}$s$^{-1}$ \cite{muon_lsc1,muon_lsc2}. \mbox{ANAIS-0} experimental layout at LSC (see Figure~\ref{fig:shielding}) consisted in a passive shielding made of 10\,cm archaeological lead plus 20\,cm low activity lead, all enclosed in a PVC box tightly closed and continuously flushed with boil-off nitrogen, and active vetoes mounted on top of the shielding to reject coincident events in \mbox{ANAIS-0} module.

\begin {figure}[ht]
\includegraphics[width=0.45\textwidth]{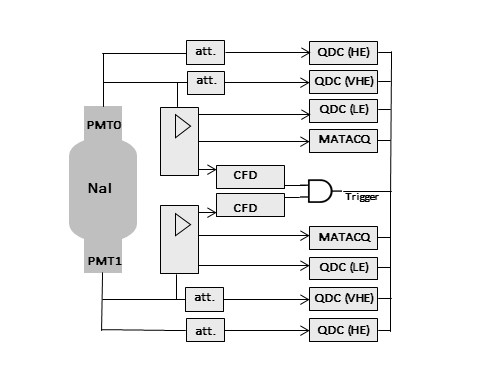}
\centering \caption{\it Simplified electronic chain diagram. Each PMT charge output signal is divided into a trigger signal, a signal going to the digitizer (MATACQ card in this case), and three signals differently attenuated and fed into QDC module channels to be integrated. Triggering is done by a logical AND of the two PMT signals enabling digitization and conversion of both separately.} \label{fig:electronics}
\end {figure}

Concerning the data readout, a simplified diagram of the electronic chain is shown in Figure~\ref{fig:electronics}. Each PMT charge output signal is separately processed: it is divided into a trigger signal, a conveniently amplified signal going to the digitizer, and three signals differently amplified or even attenuated and fed into QDC (charge-to-digital converter) module channels to be integrated in a 1\,${\mu}$s window. Triggering is done by the coincidence (logical AND) of the two PMT signals at photoelectron level in a 200\,ns window enabling digitization and conversion of the two signals. The building of the spectra is done by software (off-line) by adding the signals from both PMTs. Energy spectra are obtained from the QDCs in three different gain windows chosen to match a low, a high, and a very high energy ranges: the first one covers from threshold up to 200\,keV, the second up to about 6\,MeV, and the latter up to about 40\,MeV. In this work only the low energy range will be used. In order to obtain all the information from the shape of the pulse, each PMT signal is separately digitized.

In between the data taking corresponding to sets A and B (see Table~\ref{allsetupstab}), the electronic chain was upgraded to a combination of VME and NIM electronics, the pulse digitizer was changed from a Tektronix scope (TDS5034) to a MATACQ (CAEN~V1729) card, and a new acquisition and analysis software was developed~\cite{tfmMA}. One of the most important pieces of the electronic chain is the digitizer. The Tektronix scope presents a bandwidth of 350\,MHz, a sampling rate up to 5\,GS/s with 8 bits of resolution and adjustable dynamic range, and record length from 500 to 8$\times$10$^{6}$. A choice of 0.8\,ns/point sampling was done for the \mbox{ANAIS-0} data set~A. The MATACQ card presents a bandwidth of 300\,MHz, a sampling rate of 2\,GS/s corresponding to 0.5\,ns/point, 2520 record length, and a dynamic range of 1\,V with 12 bits of resolution.

A detailed background study of \mbox{ANAIS-0} module in the different configurations tested has been published in~\cite{ANAISbkg}. The most significant contribution to the background at low and high energy comes from $^{40}$K distributed in the bulk of the sodium iodide crystal corresponding to a specific activity of 12.7\,$\pm$\,0.5\,mBq/kg~\cite{anais40K}, too high to allow fulfilling the background goals of ANAIS experiment. However, this contamination has shown to be very useful for the characterization of the detector response in the very low energy regime, providing a population of bulk scintillation events at 3.2\,keV tagged by the coincidence with a high energy gamma (see section~\ref{third}).

\section{Energy calibration of \mbox{ANAIS-0} at very low energy}
\label{third}

\begin {figure}[ht]
\subfigure[]{\includegraphics[height=0.17\textheight]{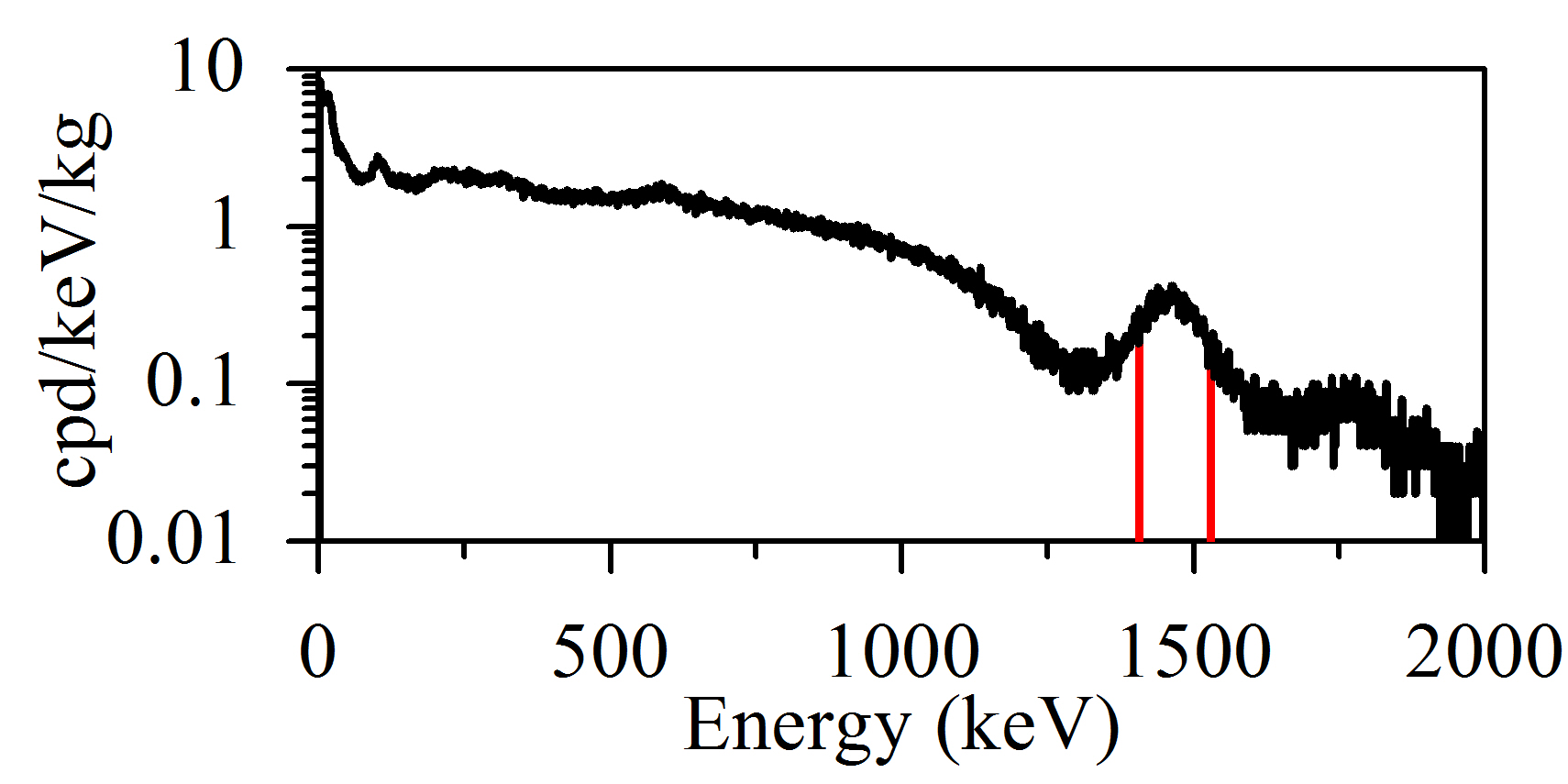}}
\subfigure[]{\includegraphics[height=0.17\textheight]{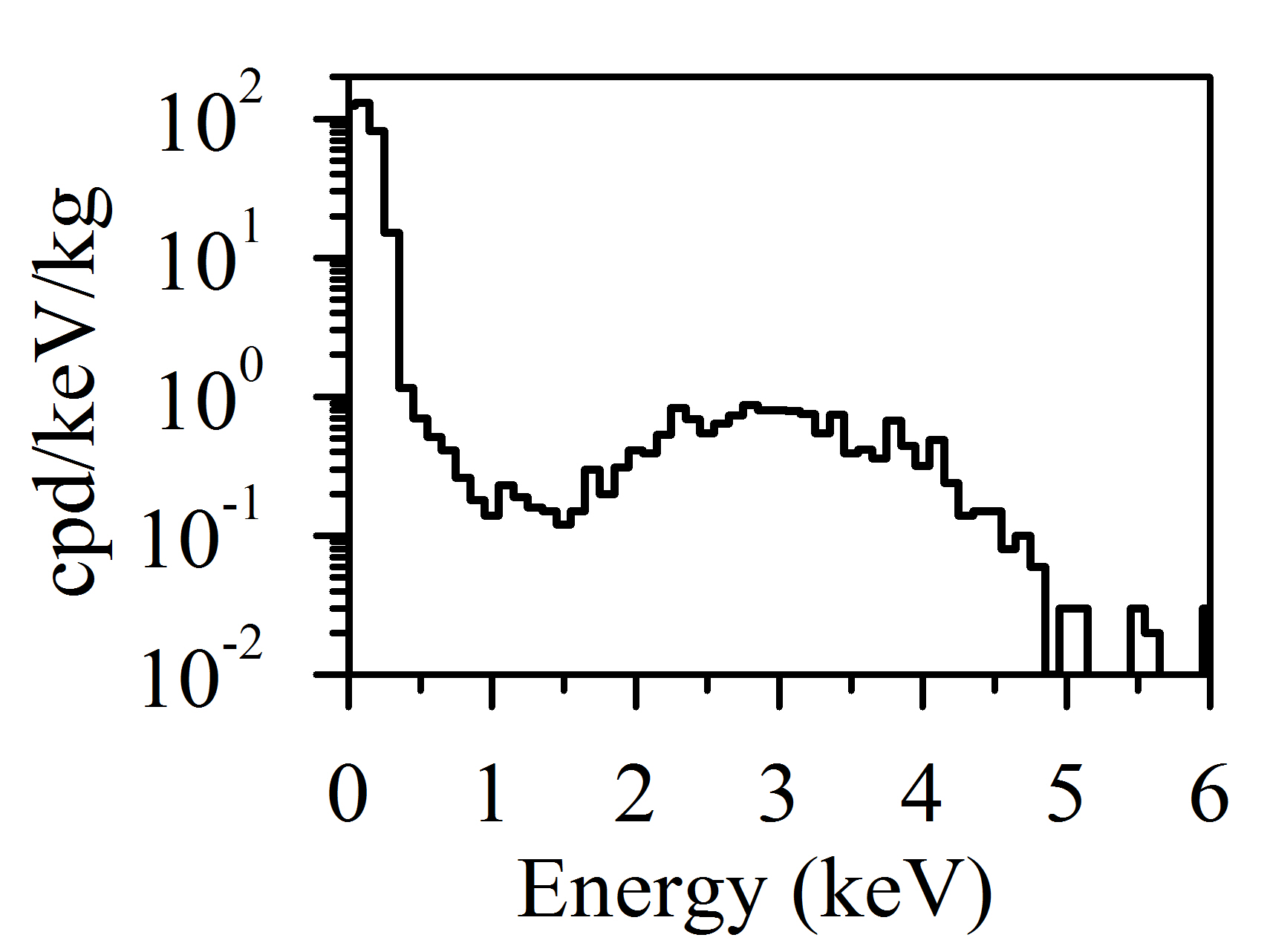}}
\centering \caption{\it (a) High energy spectrum of the 10.7\,kg detector showing in red the $1\,\sigma$ window around the 1460.8\,keV line used for the coincidence. (b) Energy distribution of the \mbox{ANAIS-0} events selected by the coincidence with the high energy window in the 10.7\,kg crystal, shown in (a).}
\label{fig:40K}
\end {figure}

The dark matter signal in NaI is expected below 10\,keVee; hence, a very good knowledge of the detector response at very low energies is mandatory, requiring powerful noise rejection protocols and reliable low energy calibration. For both purposes, reference populations of scintillation events at low energy are required. To allow that very low energy X and gamma ray emissions from external radioactive sources reach the crystal, a calibration window was included in the design of the \mbox{ANAIS-0} module encapsulation (see Figure~\ref{fig:anais0}). The external gamma calibration sources used are: $^{55}$Fe (6.0\,keV\footnote{\label{xr} X-ray energies averages have been calculated according to the relative intensities given in~\cite{nucleide}, as NaI(Tl) resolution prevents from distinguishing them individually.}), $^{109}$Cd (22.6\,keV$^{1}$ and 88.0\,keV), $^{57}$Co (6.4\,keV$^{1}$, 14.4\,keV and 122.1\,keV), and $^{133}$Ba (31.7\,keV$^{1}$ and 81.0\,keV)~\cite{nucleide}. However, the less the energy of the radiation, the less representative is the event for a bulk energy deposition being eventually affected by superficial effects in the light emission and collection. Then, in order to dispose of a population of events in the bulk at very low energy, the $^{40}$K internal contamination is very useful: 3.2\,keV energy depositions are obtained after K-shell EC in $^{40}$K in those cases the 1460.8\,keV gamma escapes completely from the detector active volume; a fraction of such events can be effectively tagged by selecting coincidences with a 1460.8\,keV energy deposition in a second detector. A dedicated set-up was specially designed and operated at LSC with such a goal: \mbox{ANAIS-0} module and a second large NaI(Tl) detector (10.7\,kg) named as prototype (PIII), and described in~\cite{anais40K,tesisClara} were placed inside the same shielding, at a distance of 13\,mm between the corresponding copper housings. This measurement was done right after the measurements corresponding to data set~B, under similar experimental conditions: in particular, same light collection efficiency than that corresponding to data set~B was checked (see section~5). This set-up allowed to determine the $^{40}$K bulk content in both crystals and to select a population of 3.2\,keV events~\cite{anais40K}. In Figure~\ref{fig:40K}.a and b are shown, respectively, the high energy spectrum of PIII and the \mbox{ANAIS-0} low energy events selected by the coincidence with the 1460.8\,keV gamma line in PIII in a 1$\,\sigma$ window, shown in red in Figure~\ref{fig:40K}.a. This event population will be used for energy calibration in this section, and for other purposes in next sections.

\begin {figure}[ht]
\subfigure[]{\includegraphics[width=0.28\textwidth]{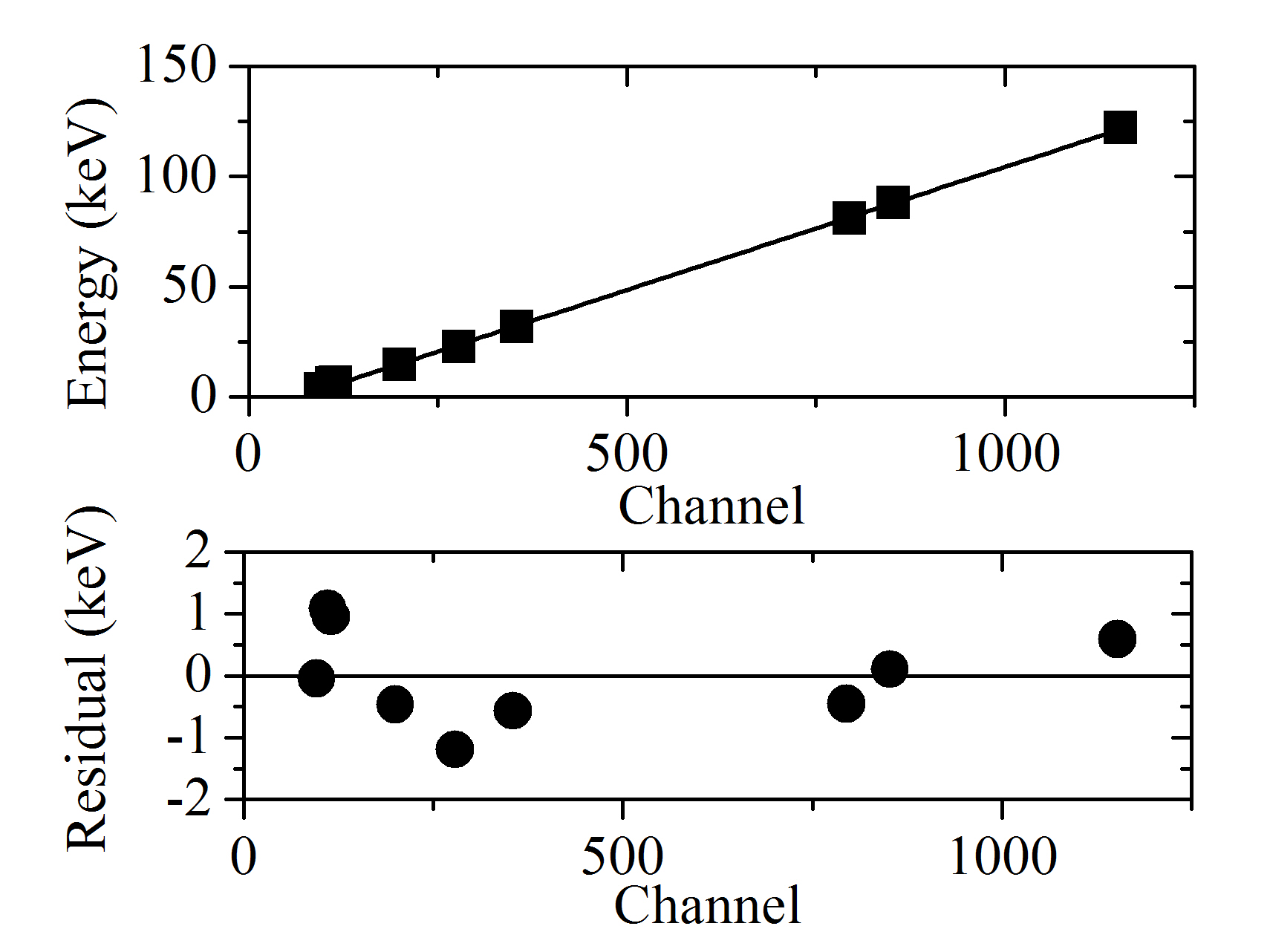}}
\subfigure[]{\includegraphics[width=0.28\textwidth]{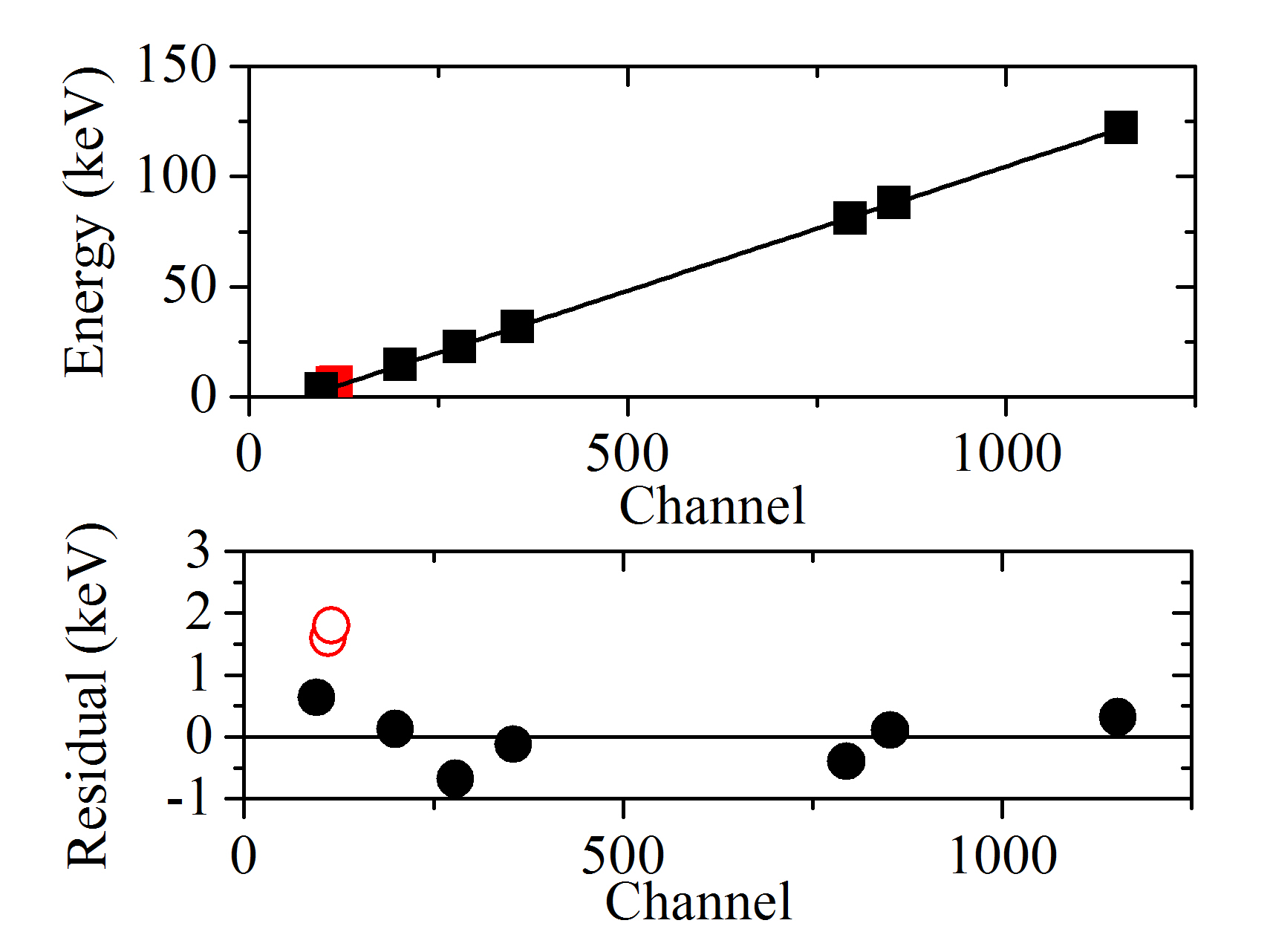}}
\subfigure[] {\includegraphics[width=0.28\textwidth]{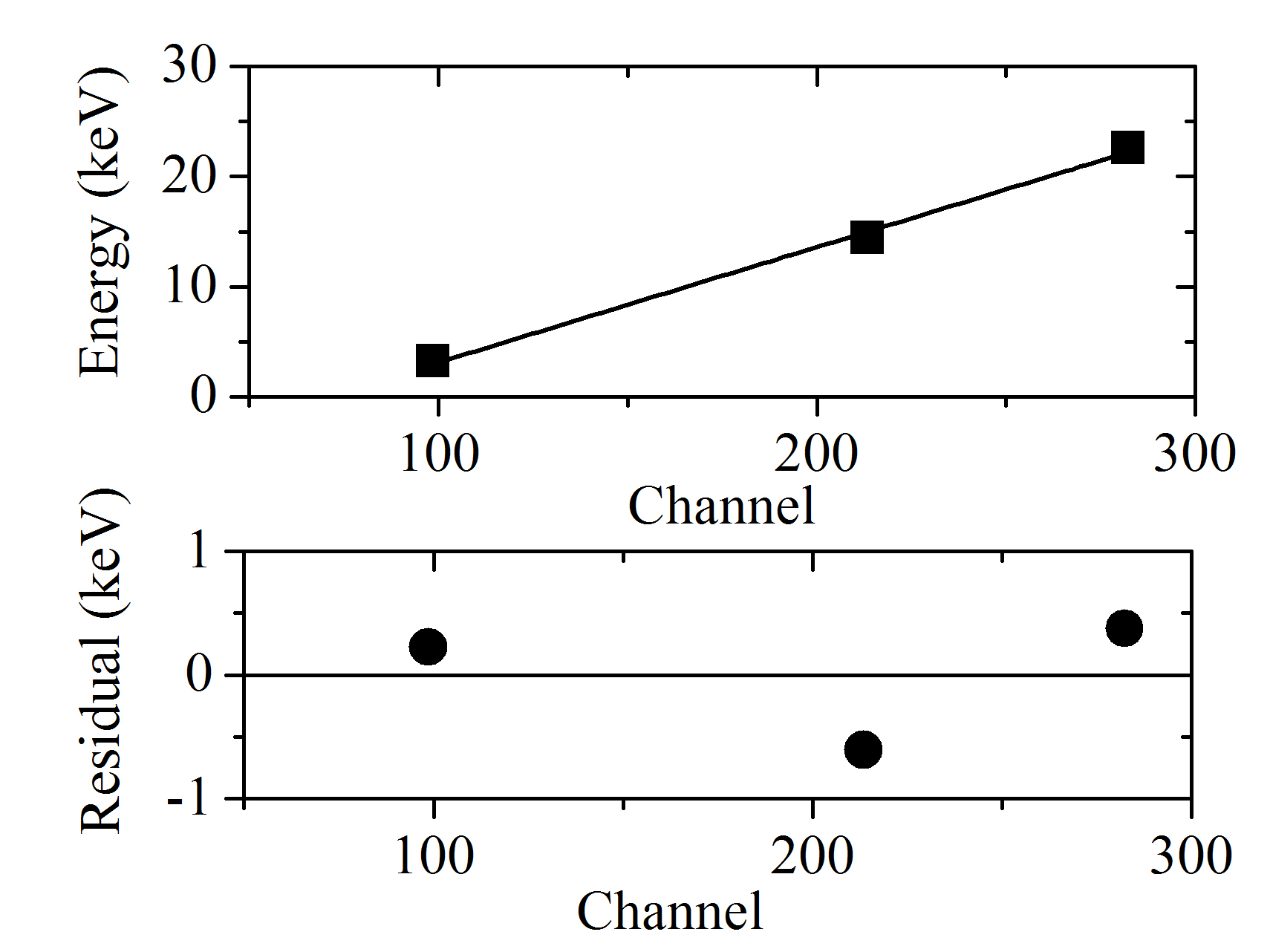}}
\centering \caption{\it (a) Energy versus QDC channel linear fit and corresponding residuals using all the available lines below 150\,keV. 6.0\,keV and 6.4\,keV lines from external origin could be strongly affected by superficial effects, and are not used in the fit shown in (b); they are marked in red. (c) Calibration chosen for the low energy region in data set A using only 3.2\,keV, 14.4\,keV and 22.6\,keV lines trying to minimize possible non-linear effects in the detector response in the dark matter relevant region of interest.}
\label{fig:linearity}
\end {figure}

One of the most important issues for ANAIS is to guarantee a good energy calibration at the lowest energies, down to the threshold. A linear fit energy vs QDC channel using all the available lines below 150\,keV during the $^{40}$K dedicated set-up is shown in Figure~\ref{fig:linearity}.a together with the corresponding residuals. It has to be remarked that residuals at 6.4 and 6.0\,keV are much larger (and both positive) considering them in percent than those corresponding to other lines. Non-proportionality in the scintillation yield with respect to the deposited energy has been observed in inorganic scintillators~\cite{birks,Dorenbos}. In particular, in NaI(Tl) scintillators some non-linear effects are expected at the K-shell Iodine binding energy (33.2\,keV). This fact has been observed experimentally~\cite{Dorenbos,lin2,lin3,Requicha,Tojo2}, but at a level of only 5\% in the amplitude response. Similarly, at around 5\,keV (L-shell Iodine binding energy) the same effect is expected, as it has been observed in Refs.~\cite{Tojo2,Khodyuk}. Nevertheless, the reported non-linearity cannot explain such a large effect as that seen in Figure~\ref{fig:linearity}.b: 6.0\,keV line appears with an effective energy of 4.2\,keV and 6.4\,keV line with 4.8\,keV. However, this effect could be due to the very low penetration depth of that radiation in the crystal that could reduce the light collected with respect to the same energy deposition for bulk events, as those from $^{40}$K. It is worth mentioning that the mean free path of a 6\,keV photon in NaI(Tl) is 5.1\,$\mu$m, whereas for a 15\,keV photon is 0.06\,mm. This hypothesis is supported by similar results obtained by several groups working with large size NaI(Tl) detectors and aiming at the detection of dark matter, that have observed good linearity at low energies, but reductions in the response to 6.0\,keV energy X-rays from an external source that were attributed to superficial effects~\cite{NuovoCim,gerbier99}. In Ref.~\cite{gerbier99} a Compton scattering experiment was designed in order to obtain a volume-distributed low-energy population whose results supported such assumption.

Hence, we removed both lines from the calibration procedure, obtaining the fit shown in Figure~\ref{fig:linearity}.b. Finally, as the dark matter region of interest in NaI(Tl) scintillators is below 10\,keVee, it was decided to calibrate the low energy range using only 3.2\,keV, 14.4\,keV and 22.6\,keV lines trying to minimize possible non-linear effects in the detector response. In Figure~\ref{fig:linearity}.c are shown the linear fit and residuals corresponding to the low energy calibration for data set A (similar calibration has been applied to data set B). Then, 0.5\,keV can be taken as a reasonable estimate of the systematic error of our low energy calibration in the dark matter region of interest.

\section{Trigger efficiency estimate}
\label{sec3,5}

Lowering the threshold as much as possible is mandatory in any experiment devoted to the direct search for dark matter. The lowest achievable threshold in a scintillation experiment implies, as first step, to trigger at single photoelectron (phe) level in each PMT signal. We have studied the trigger level in \mbox{ANAIS-0} by two different methods: first, the distribution of amplitude of the Single Electron Response (SER) is compared to the amplitude of a single peak (at the trigger position), and second, the 3.2 keV events population selected by the coincidence has been profited to study how many of them trigger effectively our acquisition.

\begin {figure}[ht]
\includegraphics[width=0.45\textwidth]{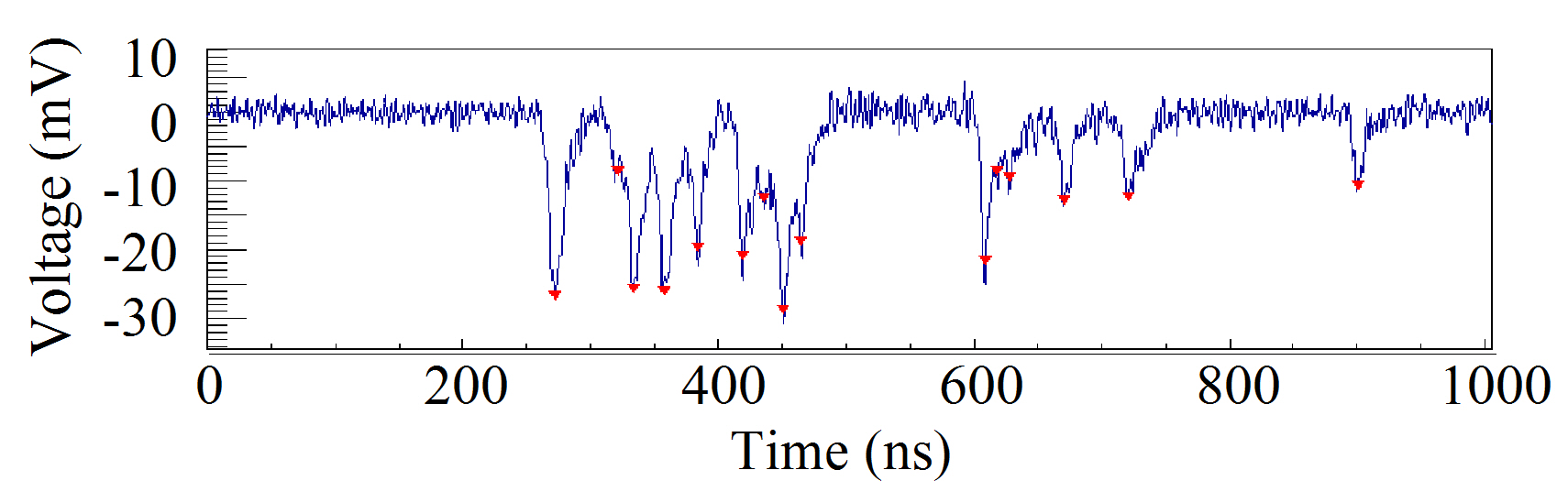}
\centering \caption{\it Low energy pulse (from the population of 3.2\,keV events after $^{40}$K decay, selected by coincidence) showing peaks identified by the peak search algorithm. They are counted independently for each PMT signal and their positions saved. The pulse shown corresponds to data set~B (data taken with the MATACQ).}
\label{fig:n}
\end {figure}

Profiting from the high sampling rate of the digitized data, the discrete arrival of the scintillation photons to the PMT photocathode can be distinguished at low energies. An algorithm that determines the number of peaks in the pulse has been developed. It is based on the TSpectrum ROOT class and the Search method~\cite{Morhac}. Peaks are considered gaussian with a minimum height and width, which are selected specifically for every data set, as the SER depends on the PMT model. By choosing these parameters according to the SER of the PMT, peaks are found, counted and their positions saved. In Figure~\ref{fig:n}, a low energy pulse is shown and the peaks identified by the algorithm are marked. It is worth noting that this algorithm can only be applied to events with a low number of phe, otherwise the number of peaks would underestimate the number of phe, since the probability of two phe overlapped is not negligible.

\begin {figure}[ht]
\subfigure[]{\includegraphics[width=0.43\textwidth]{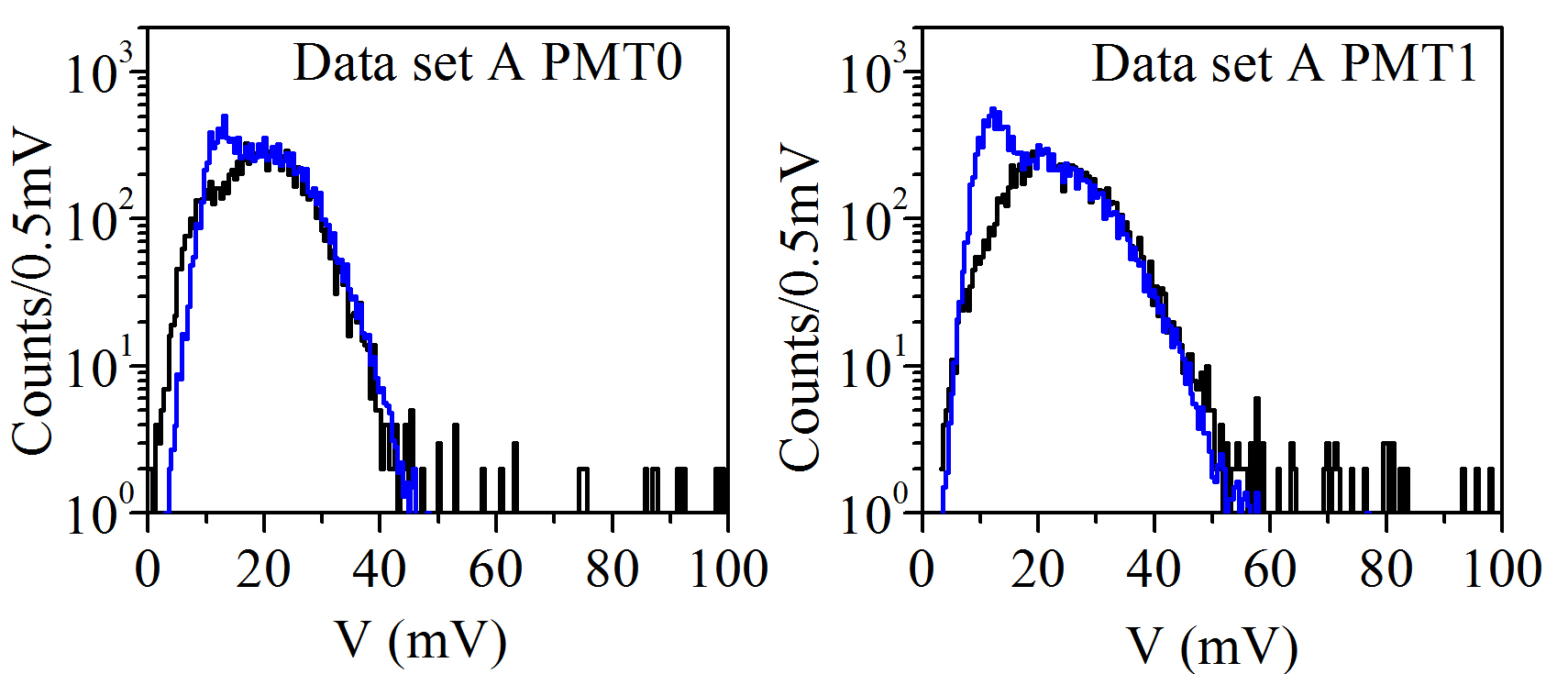}}
\subfigure[]{\includegraphics[width=0.43\textwidth]{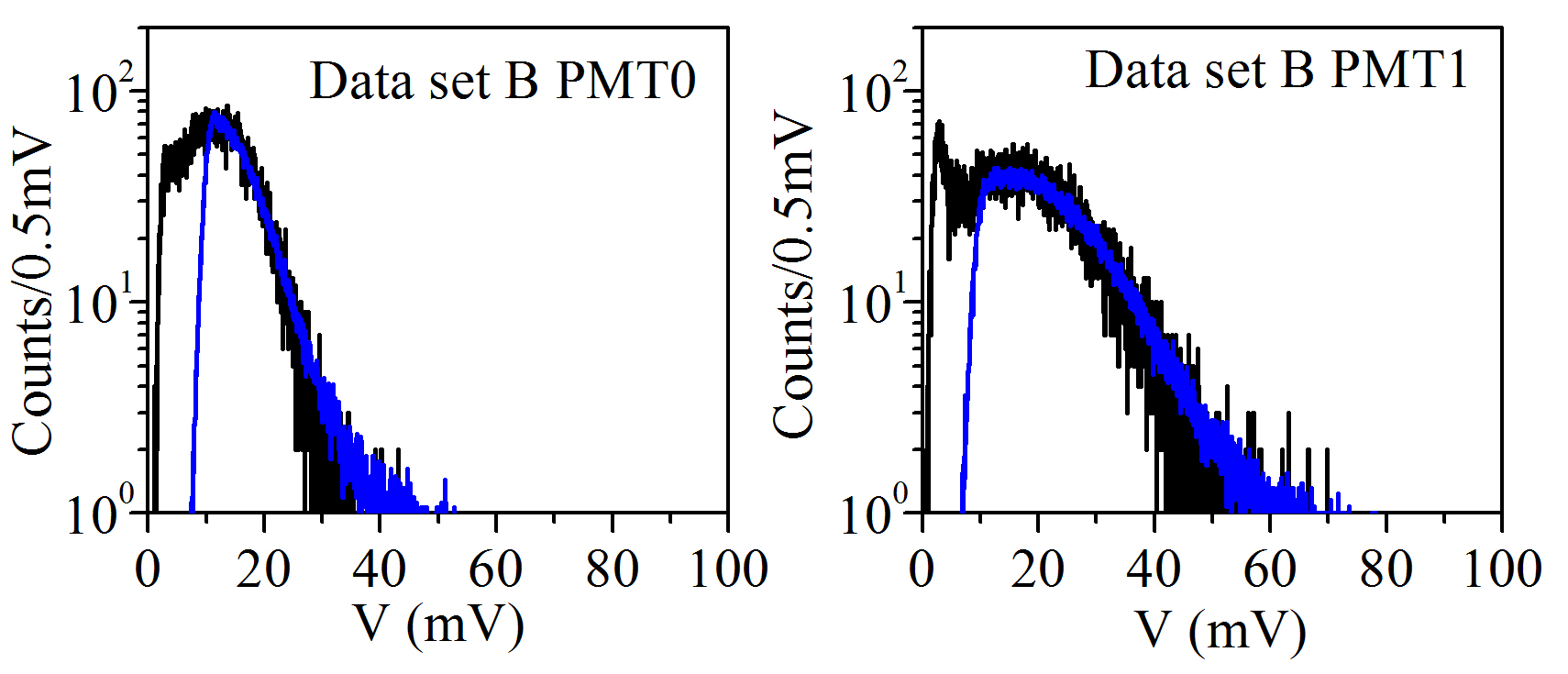}}
\centering \caption{\it SER (black) and single peak amplitude (blue) distribution for data sets A and B. Different PMT models have been used in each data set.}
\label{fig:ampphe}
\end {figure}

The SER parameters (amplitude, area, etc.) are determined from a population of single photoelectron peaks built by selecting the last identified peak by this algorithm in pulses having a low number of peaks. As far as only peaks found in the last part of the pulse are used to build the SER, bias related to effective trigger efficiency is avoided. Figure~\ref{fig:ampphe} compares the amplitude distribution of the SER and that corresponding to a single peak (responsible of the trigger) for pulses having only one peak per PMT. The SER amplitude spectra have been gaussian fitted: mean values and standard deviations are shown in Table~\ref{tab:ampphe}.
Full trigger at photoelectron level has been achieved in data set A, as it can be observed in Figure~\ref{fig:ampphe}: single peak amplitude distribution follows quite well the SER distribution plus a baseline noise component also triggering the acquisition; on the other hand, an effective trigger threshold at 12\,mV is observed in data set B. In this case, by comparing such effective trigger threshold with the mean value and standard deviation of the SER amplitude distribution gaussian fit, the percent of the SER distribution effectively triggering is calculated (see Table~\ref{tab:ampphe}). As a conclusion, full trigger at phe level was achieved in data set~A, whereas trigger at or above 50\% phe level can be reported for data set~B.

\begin{table}[ht]
\begin{center}
\fboxrule=0cm \fbox{
\begin{tabular}[]{lllll}
\toprule
Data set	& Signal		&	$\mu_{SER}$ 	& $\sigma_{SER}$ & Eff.\\
\cmidrule{3-5}
				& 				& mV		&	mV				& \%\\
\cmidrule{1-5}
\cmidrule{1-5}
A	&	PMT\,0	& $20\pm1$	& $7.3\pm0.1$&  100\\
	&	PMT\,1	& $23\pm1$	& $8.0\pm0.1$&  100\\
B	&	PMT\,0	& $12\pm1$	& $5.9\pm0.1$&  $50$\\
	&	PMT\,1	& $16\pm1$	& $11\pm1$& 	$65$\\
\bottomrule
\end{tabular}
}\caption{\it Mean photoelectron amplitude ($\mu_{SER}$), standard deviation ($\sigma_{SER}$), together with the percent of the photoelectron distribution effectively triggering every PMT signal channel (see text for details). Data are shown for data sets A and B. If the existence of an unphysical region of negative amplitudes is taken into account, recalculated triggering efficiencies are 60\% (data set A) and 70\% (data set B).}
\label{tab:ampphe}
\end{center}
\end{table}

In the case of data set A, for which a lower light collection was obtained ($\sim$2.7\,phe/keV/PMT~\cite{tesisClara}), the influence of the coincidence window width on the trigger efficiency cannot be neglected. Monte Carlo simulation of the phe arrival at every PMT for a given energy deposition, supposing for bulk NaI scintillation events phe are distributed following an exponential decay with $\tau$\,=\,230\,ns and the reported light collection efficiencies. The results for a 200\,ns coincidence window indicate an efficiency of 97\% in the 2-3\,keV energy bin and 100\% above 3 keV. In the case of data set B, with a light collection of $\sim$3.7\,phe/keV/PMT~\cite{tesisClara}, the coincidence window width effect on the trigger efficiency is negligible above 2\,keV.

\begin {figure}[ht]
\includegraphics[height=0.15\textheight]{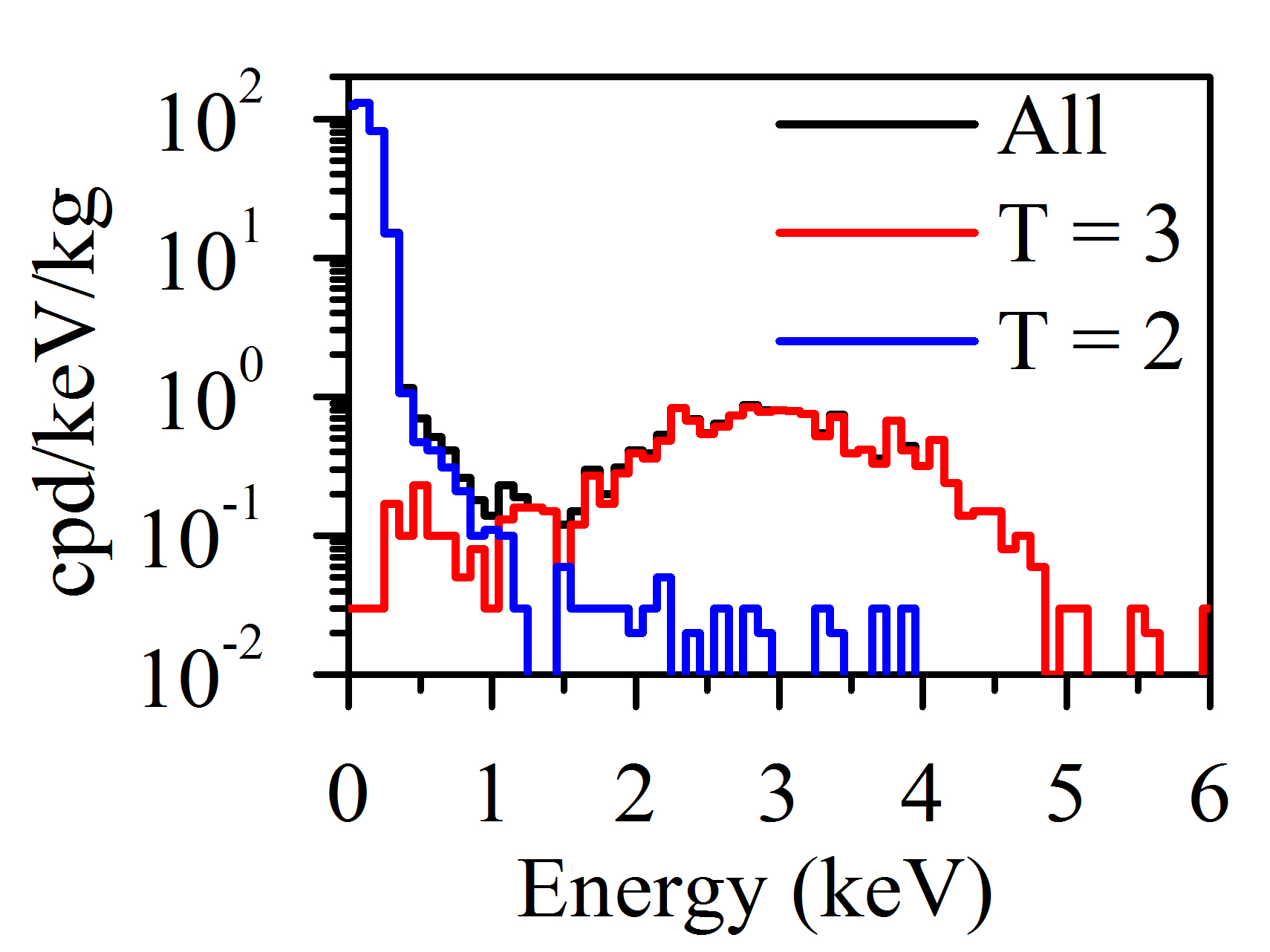}
\centering \caption{\it Low energy spectra of \mbox{ANAIS-0} selected by the coincidence with the 1460.8\,keV gamma line in the PIII considering T variable. All events (in black), events having T\,=\,3 (in red) and T\,=\,2 (in blue) are shown. See text for explanation of the variable T.}
\label{fig:40KPIV_malu}
\end {figure}

In order to quantify the trigger efficiency at threshold in a direct way, we profit from the availability of the selected population of 3.2\,keV presented in the previous section. The trigger configuration of every event is defined in a 1\,$\mu$s coincidence time window and described with the T variable: T\,=\,1, if only \mbox{ANAIS-0} triggered; T\,=\,2, if only PIII triggered; and T\,=\,3 if both detectors triggered. In Figure~\ref{fig:40KPIV_malu} spectra of the \mbox{ANAIS-0} module in coincidence with the 1460.8\,keV line in the PIII are shown.  No events with T\,=\,1 are found in this population because the imposition of the coincidence implies that PIII always triggers. Events in the peak at 3.2\,keV (those above 1.5\,keV in Figure~\ref{fig:40KPIV_malu}) triggered with almost full efficiency (98\%), supporting a very high trigger efficiency at or even below 2\,keVee. We derive this efficiency as the fraction of events above 1.5\,keV and below 6\,keV triggering the ANAIS-0 module from the total of events in the same energy region selected by the high energy window (around $^{40}$K gamma) in the PIII module. In conclusion, very high trigger efficiency above 2\,keVee has been achieved with \mbox{ANAIS-0} prototype and full trigger at photoelectron level has shown to be achievable, fulfilling the goal of ANAIS experiment.

\section{Bulk NaI scintillation low energy events selection}
\label{forth}

\begin {figure}[ht]
\includegraphics[width=0.45\textwidth]{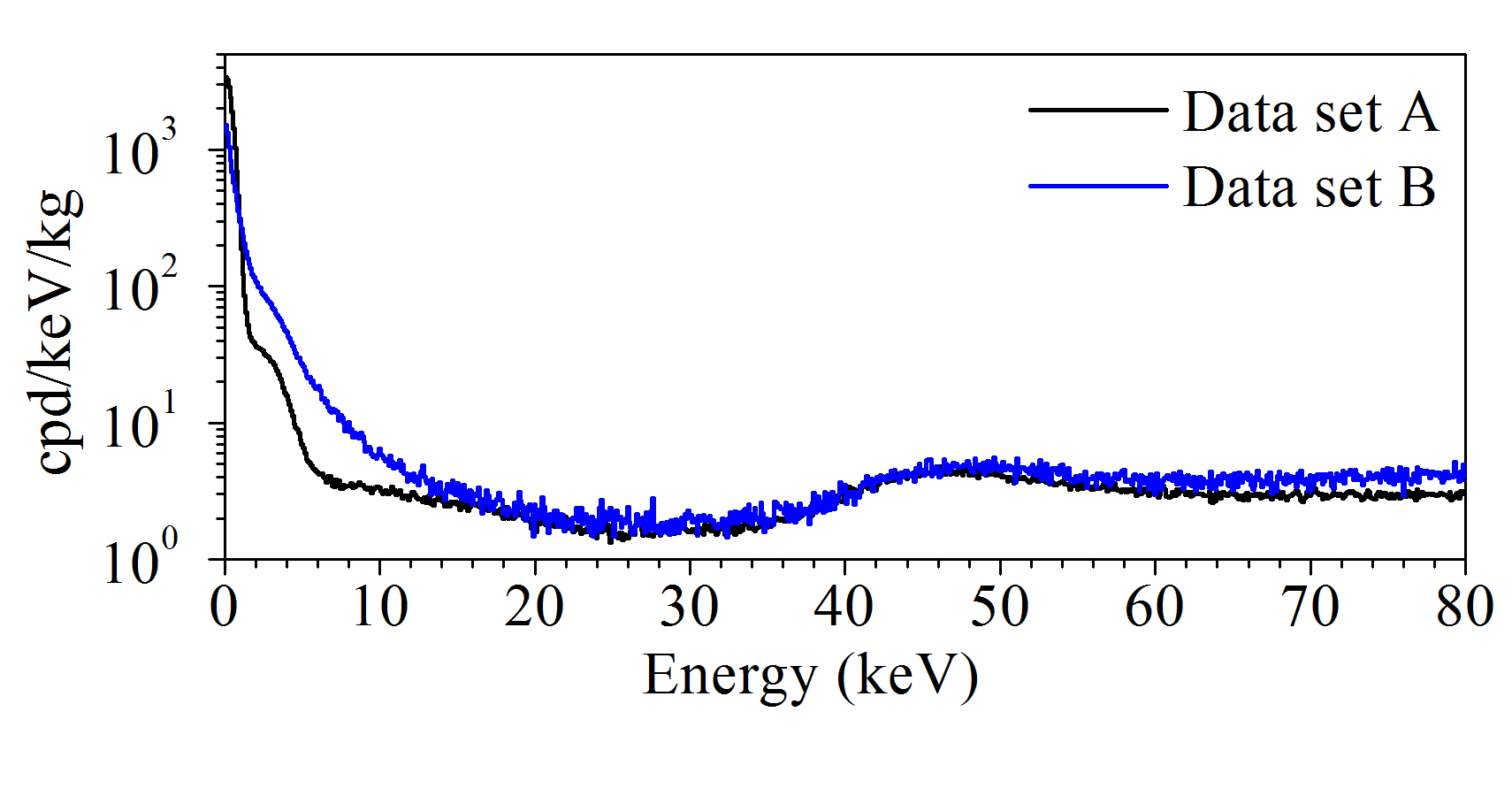}
\centering \caption{\it Spectra of the raw events for data set A (black) and B (blue). The high increase of events rate below 20\,keV, clearly dependent on the PMT model, is attributed to PMT origin events.}
\label{fig:f1}
\end {figure}

Below 10\,keVee, raw data from NaI scintillator detectors are dominated by events other than those from scintillation in the bulk, having mainly their origin in the PMTs. Strong rejection of such kind of events is required, see Figure~\ref{fig:f1}, in order to reduce the effective energy threshold down to 2\,keVee. Dark matter particles are expected to interact in our detector by elastic scattering off Na and I nuclei. Hence, the energy is deposited through the interaction of the corresponding recoiling nucleus in the crystal. Dark matter events are expected to be very similar to those produced by neutrons, and to share some features with those having beta/gamma origin; the last two can be produced using calibration sources in order to have reference populations that allow a good NaI(Tl) bulk scintillation event characterization. All the events populations non attributable to dark matter interactions should be conveniently filtered. This implies to remove from raw data those events correlated with muon interactions in the vetoes, coincidences between two or more modules, events accumulated in short time periods, and events showing anomalous scintillation time constants in the pulse shape.

The goal is to find a compromise between a high acceptance of bulk NaI scintillation events and low contribution of other spurious events, not rejected by the filtering. This filtering procedure implies in some cases an effective reduction in the acquisition live time, while in others the efficiency of the selection filter to preserve the bulk scintillation events in NaI(Tl) active volume has to be estimated by studying reference populations, specially at very low energies. $^{57}$Co, $^{109}$Cd, and $^{133}$Ba calibration data and the 3.2\,keV events following $^{40}$K decay in the bulk have been used for that purpose, whereas $^{252}$Cf calibration data will be analyzed in section~6.

We describe below in detail the filtering procedure followed with \mbox{ANAIS-0} data. Table~\ref{tab:cuts} shows the efficiency factor considered, as well as the available live time, before and after the application of each filtering procedure to the two data sets considered in this work. All the filters have been applied consecutively and in the order presented in the text to every data set.

\begin{table}[ht]
\caption{\it Efficiency correction factor (Eff.) considered and live time (LT) remaining after the application of the different filters described in the text to the two data sets considered in this work. ($\ast$) In filter~3, the efficiency is 100\% above 4\,keV in both data sets, we indicate in this table the lower value of such efficiency (that corresponding to the 2-3\,keV energy bin), see text for details; in filter~4, two different acceptance criteria have been used, and in the second case the efficiency is energy dependent (see text and Figure~\ref{fig:eff} for details).}
\label{tab:cuts}
\begin{tabular}{lllll}
\hline\noalign{\smallskip}
Filter & \multicolumn{2}{l}{Data set A}& \multicolumn{2}{l}{Data set B} \\
			&Eff.	& LT (s) &Eff.	& LT (s) \\
\noalign{\smallskip}\hline\noalign{\smallskip}
Raw 		& - 	&  10649269 &-&2755985\\
1 			& 100\% 	& 10648800&100\%&2755660\\
2 			& 100\%    & 10648800 &99.8\%&2755660\\
3 			& $>$78\% ($\ast$)   & 10648800 &$>$95\% ($\ast$) &2755660\\
4			& 97.7\%/($\ast$)   & 10648800 &97.7\%/($\ast$) &2755660\\
\noalign{\smallskip}\hline
\end{tabular}
\end{table}

\begin{enumerate}

	\item \textbf{Muon related events.}
Active plastic scintillator vetoes were installed on top of the \mbox{ANAIS-0} shielding to reject the residual cosmic muon flux contribution to the background of ANAIS, and also to monitor the muon rate in the laboratory at the shielding position in order to evaluate any possible seasonal variation. A good comprehension of the muon related events in the ANAIS experiment is required because the annual modulation in the muon rate is well known~\cite{muonsborexino,macro,lvd}, and it should be discarded as responsible of any modulation observed in the very low energy events rate. This issue has been discussed in the frame of the DAMA/LIBRA experiment~\cite{nygren,muons2,muons1,muons3}, and is even more important for the ANAIS experiment because residual muon flux at LSC is about one order of magnitude larger than at Gran Sasso Laboratory, given the smaller rock overburden. Muon coincident events in \mbox{ANAIS-0} have been identified using the time after the last muon event in the plastic scintillators: a counter is reset by any event triggering one (or more) of the active vetoes; the value of this counter is read and associated to every event in the NaI(Tl) detector. Events coincident in a 20\,$\mu$s window with a signal in the muon vetoes  are rejected off-line. As the \mbox{ANAIS-0} acquisition rate was $\sim$1\,Hz, and muon coincident events rate during data set A was 43.43\,$\pm$\,0.05\,cpd, neither live time nor efficiency corrections are required.

However, muons can produce other kind of events, non-coincident with the veto signals, as delayed neutrons. Furthermore, when a very high energetic particle interacts in the \mbox{ANAIS-0} module, photons emitted in the tail of the pulse, up to hundreds of ms after the pulse onset, are able to trigger again the acquisition because of the very slow NaI(Tl) scintillation evidenced in~\cite{ANAISom} and the setting of the trigger at photoelectron level. We observed a clear increase in the total acquisition rate after every very high energy deposition event (see Figure~\ref{fig:muonact}), many (but not all) of them could be identified by the coincidence with a signal in the muon veto scintillator because of the partial coverage. For that reason, in data set A all the events triggering during 0.5\,s after a high energy event (over 9\,MeVee to guarantee to be well above the usual alpha and gamma backgrounds) are rejected and the corresponding live time deducted.  Nevertheless, in data set B, because PMT signals saturated at energies much below 9\,MeVee, it was decided to reject 0.5\,s after the arrival of a muon at the plastic vetoes (conservative approach). The same criterion was also applied to data set A to verify its compatibility. Spectra of events rejected in both data sets are shown in Figure~\ref{fig:LEcut3}. Rates of events rejected by this filter in the 2-20 keV region are 4.39\,cpd/kg and 8.65\,cpd/kg for data sets A and B, respectively. Main difference between both spectra is found in the 2-6\,keV region, and can be explained by considering the different characteristics of PMT models used in each data set: quantum efficiencies are different and PMT body consists of Kovar metal in data set~ A and glass in data set~B, allowing in the latter for the production of Cerenkov radiation in the PMT itself after a direct muon interaction.

\begin {figure}[ht]
\subfigure[]{\includegraphics[width=0.3\textwidth]{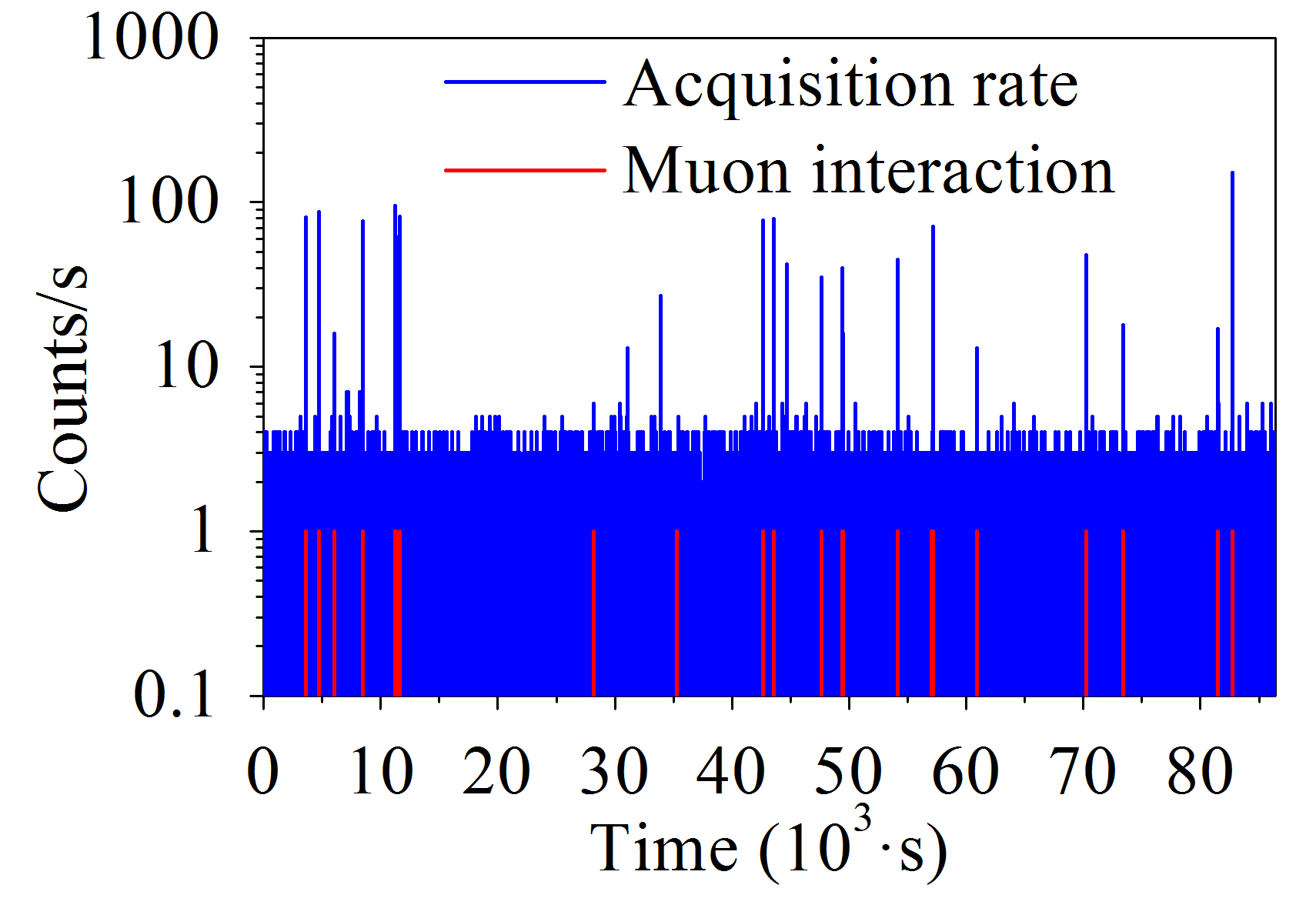}}
\subfigure[]{\includegraphics[width=0.3\textwidth]{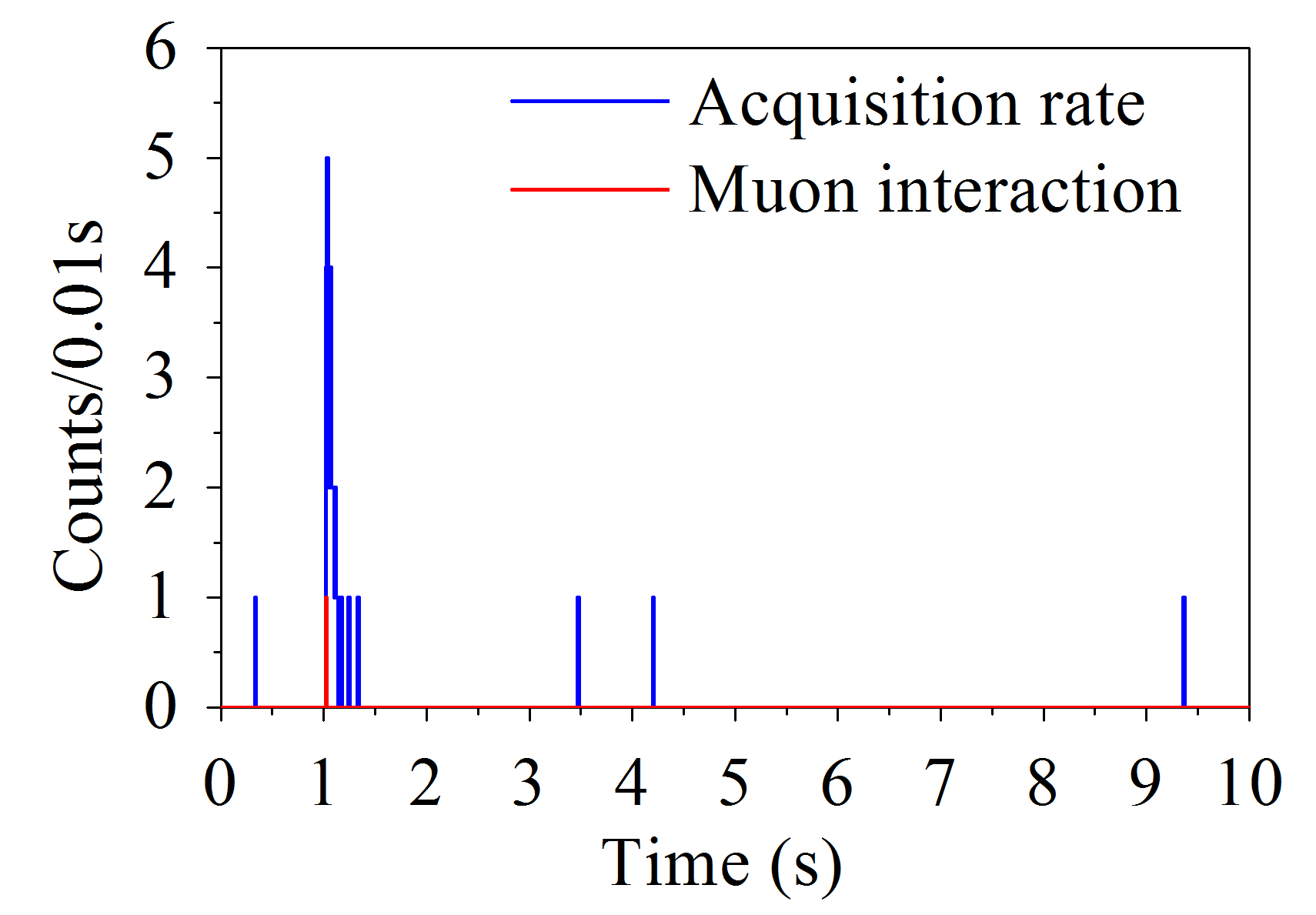}}
\centering \caption{\it \mbox{ANAIS-0} module total trigger rate (in blue) along a week (a), and 10\,s zoom (b). In red, very high energy events (above 9\,MeVee) are marked. It can be observed the clear correlation between these events (mostly attributable to muon interactions in the NaI(Tl) crystal) and the increase in the trigger rate.}
\label{fig:muonact}
\end {figure}

\begin {figure}[ht]
\includegraphics[width=0.45\textwidth]{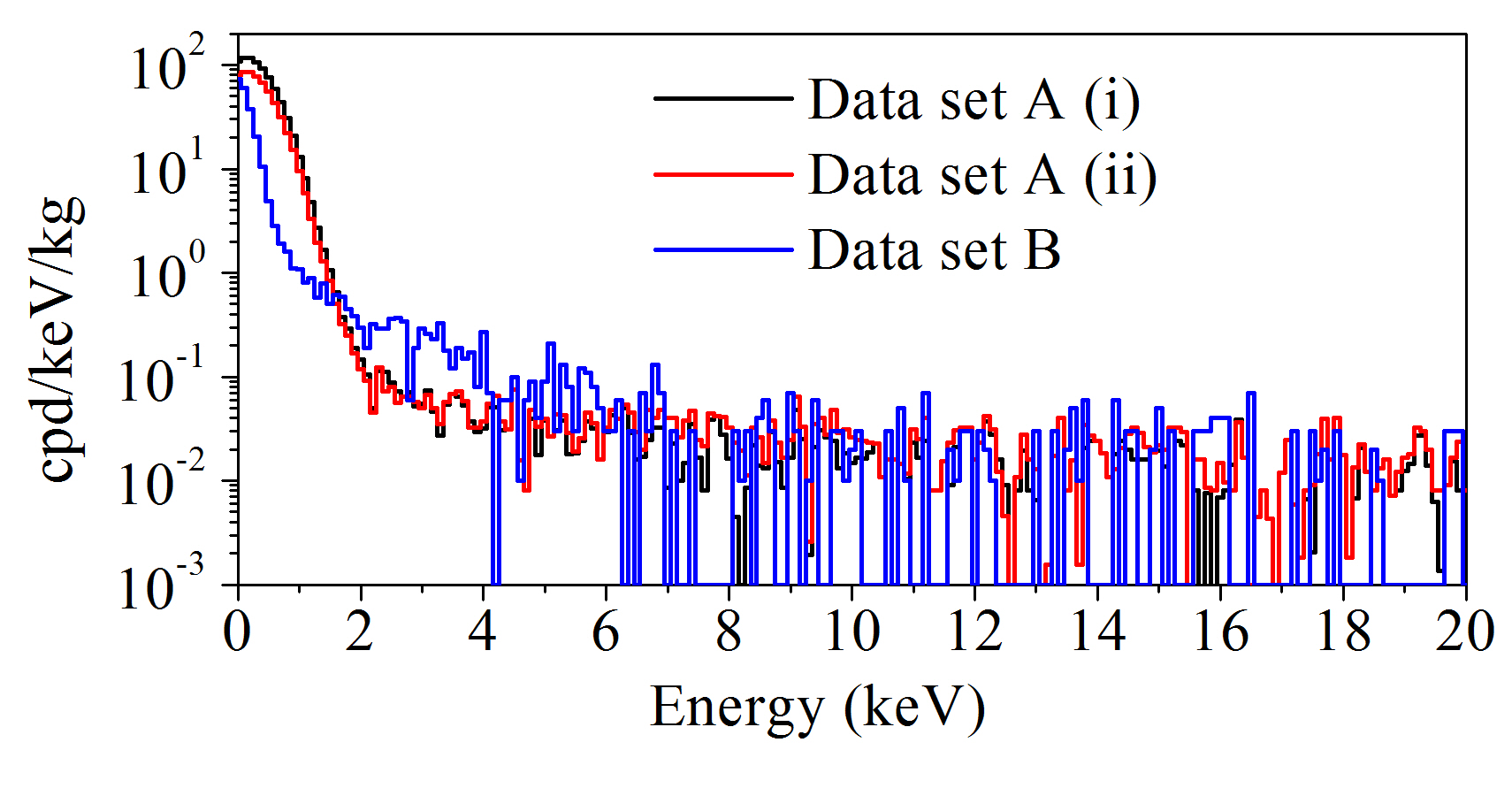}
\centering \caption{\it Low energy spectra of the muon related events rejected for data set A (black) and B (blue). Muon related events that would be rejected for data set A using the criterion applied in data set B are shown in red. See text for details.}
\label{fig:LEcut3}
\end {figure}

	\item  \textbf{Events having an anomalous baseline estimate.}
The baseline or DC-level is calculated for every event pulse by averaging the first points of the pretrigger region, clearly before the pulse onset. If a photon arrives in the pretrigger region, neither the baseline will be properly calculated, nor other related pulse parameters. These events are easily identified by their anomalous low baseline level and they will not be considered for the analysis, see Figure~\ref{fig:baseline}. They can be attributed to tails of pulses which arrived during the DAQ rearm time after a previous event, or PMT dark current photoelectrons. During data set A, the baseline was calculated with 100 points (80\,ns) whereas during data set B, after the electronic chain upgrade, with 500 points (250\,ns); hence, more events are rejected by this filter in data set B. In addition, R6956MOD PMTs (used in data set~B) present a higher dark current rate than R11065SEL PMTs (used in data set~A) leading also to reject more events by this filter. A 99.8\% of the events above 2\,keV pass the filter in data set B, and a 100\% in data set A, indicating that our filtering is not removing significantly events above our threshold (see later). These numbers can be considered as the efficiency of the cut in a conservative way. Although they represent a small percentage of the total number of events, work is in progress to recalculate the baseline for these events, and this filter could be avoided in the future.

\begin {figure}[ht]
\subfigure[]{\includegraphics[width=0.3\textwidth]{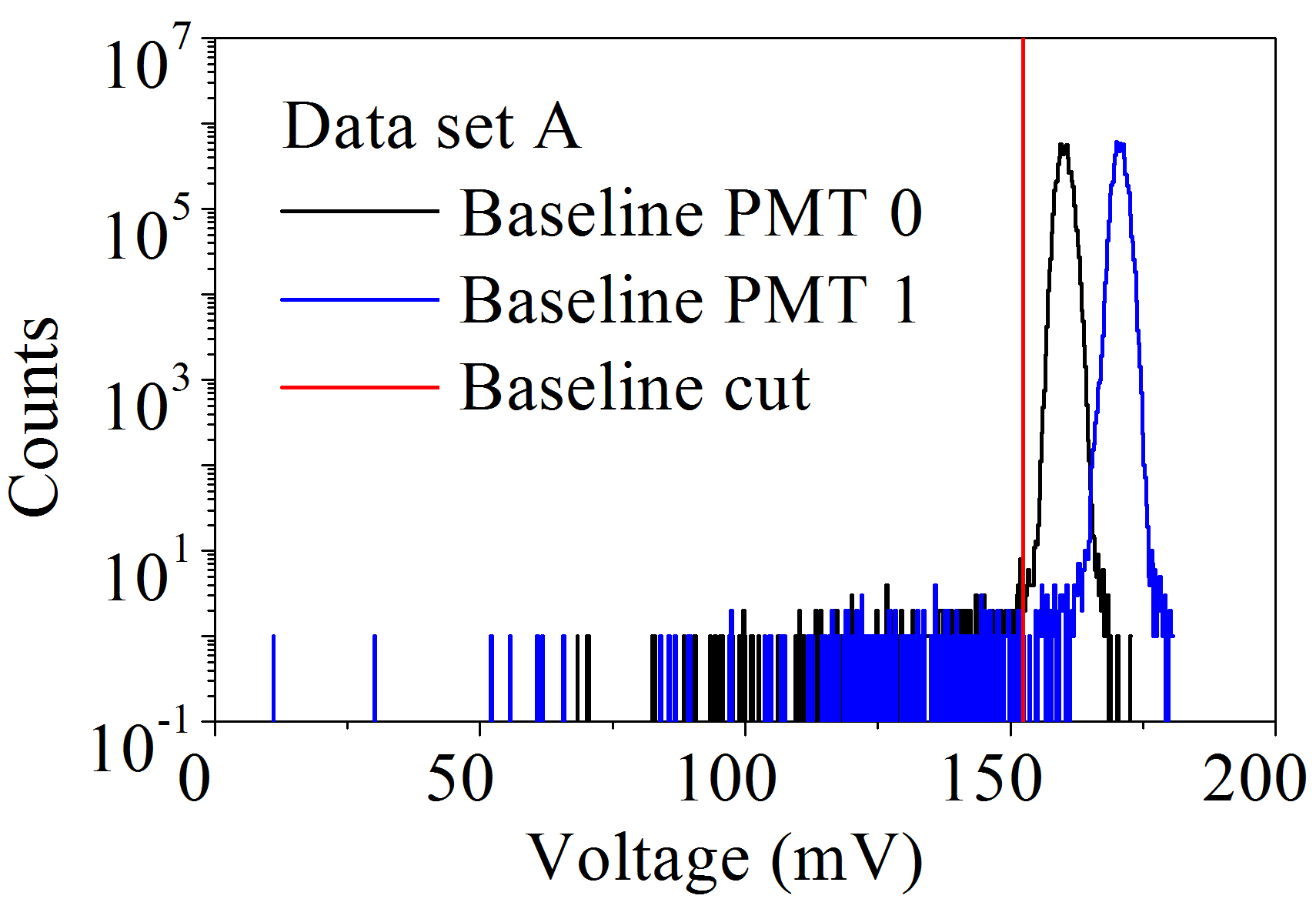}}
\subfigure[]{\includegraphics[width=0.3\textwidth]{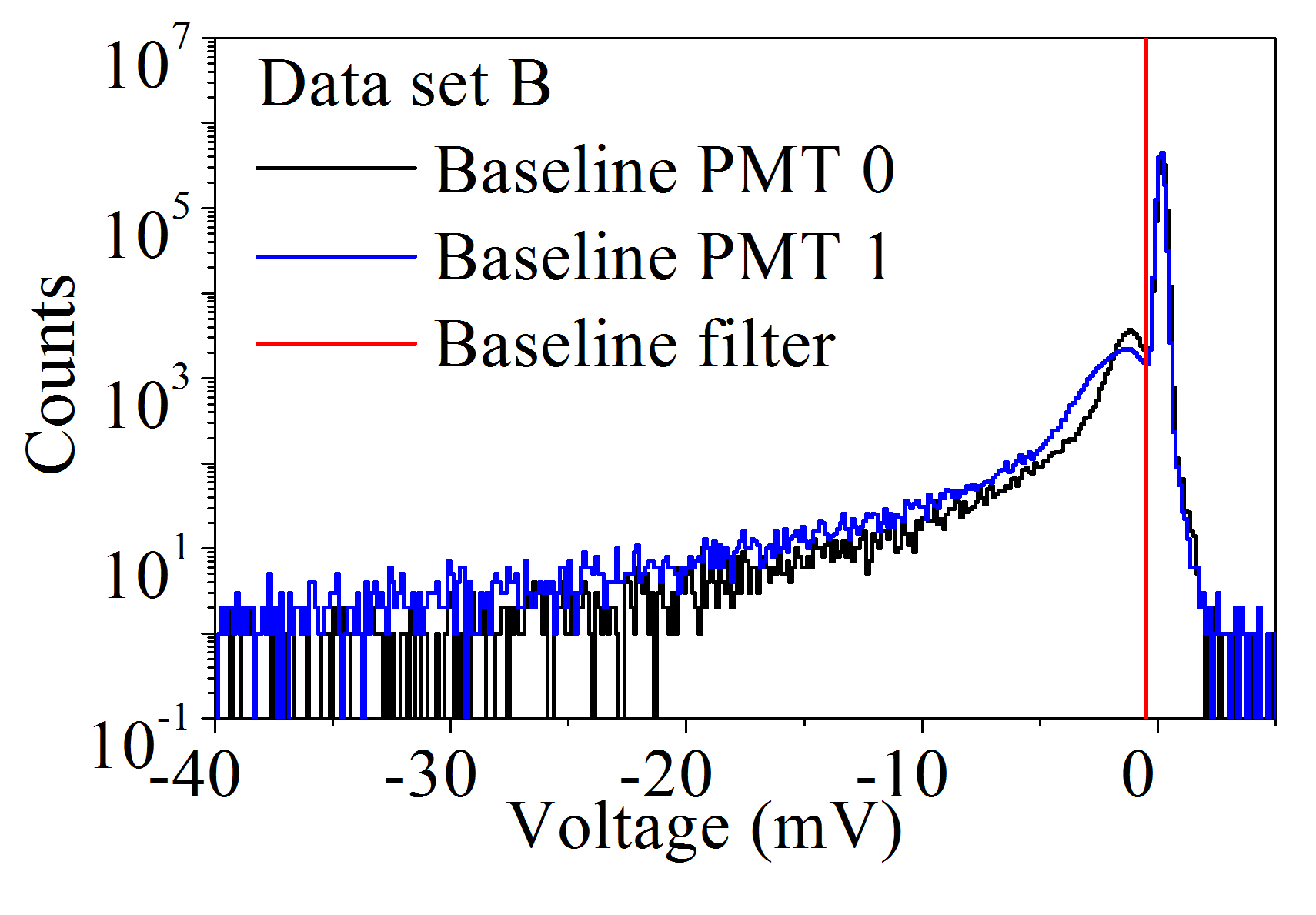}}
\centering \caption{\it Distribution of baseline parameter for PMT\,0 and PMT\,1 data corresponding to data sets A~(a) and~B~(b). Cut value applied to reject events having anomalously estimated baseline is shown in red. Data set A pulses were digitized with the scope, and data set B with the MATACQ, showing different typical baseline values.}
\label{fig:baseline}
\end {figure}

	\item \textbf{Events having a very low number of peaks.}
We reject events having $\leq$3 peaks in any of the PMTs, applying the algorithm that determines the number of peaks in the pulse described in section 4. According to the light yield measurements, $5.34\pm\,0.05$\,phe/keV in data set A and $7.38\pm\,0.07$\,phe/keV in data set B~\cite{tesisClara}, this implies an effective analysis threshold below 2\,keVee. This filter allows to reject events triggering due to a chance coincidence between uncorrelated photoelectrons in both PMTs (directly related to their respective dark currents), or events having their origin in the PMTs due to its own radioactivity (possible Cerenkov light emission in the PMT glass, for instance) that are expected to produce a signal very similar to SER, except in amplitude/area. The effect of this filter in data of a $^{57}$Co calibration is shown in Figure~\ref{fig:co57n} for data sets~A and B. This filter is mostly removing events below 2\,keVee, and 6.4 and 14.4\,keV lines are not affected by this filter at all. Same conclusion can be drawn from the $^{40}$K selected population at 3.2\,keV, as it can be seen in Figure~\ref{fig:co57n}.c. We have estimated the efficiency of this cut assuming Poisson distribution of the number of phe according to the light collection measurements. In data set A, a 78\% efficiency in the 2-3\,keVee energy region is derived, a 96\% in the 3-4\,keVee region, and a 100\% above 4\,keVee. In data set B, a 95\% efficiency in the 2-3\,keVee energy region is derived, and a 100\% acceptance above 3\,keVee for bulk scintillation events. This cut efficiency has been also estimated with the 3.2\,keV population coming from $^{40}K$ decay (described in section 3), resulting a slightly higher value than the previously reported. In the following we apply the efficiencies derived by assuming Poisson distribution of the number of peaks in the pulse, being the most conservative choice.

\begin {figure}[ht]
\subfigure[]{\includegraphics[width=0.3\textwidth]{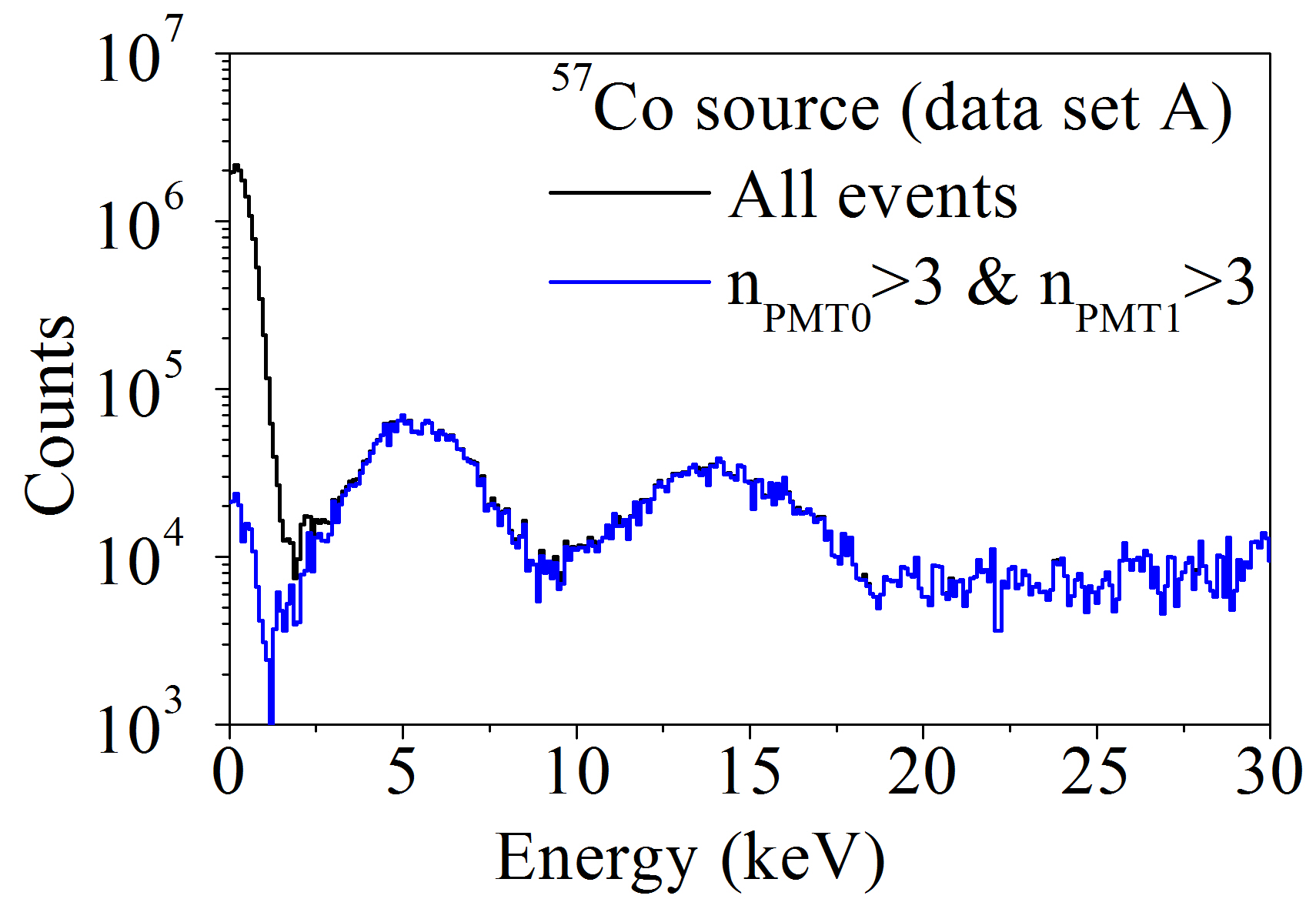}}
\subfigure[]{\includegraphics[width=0.3\textwidth]{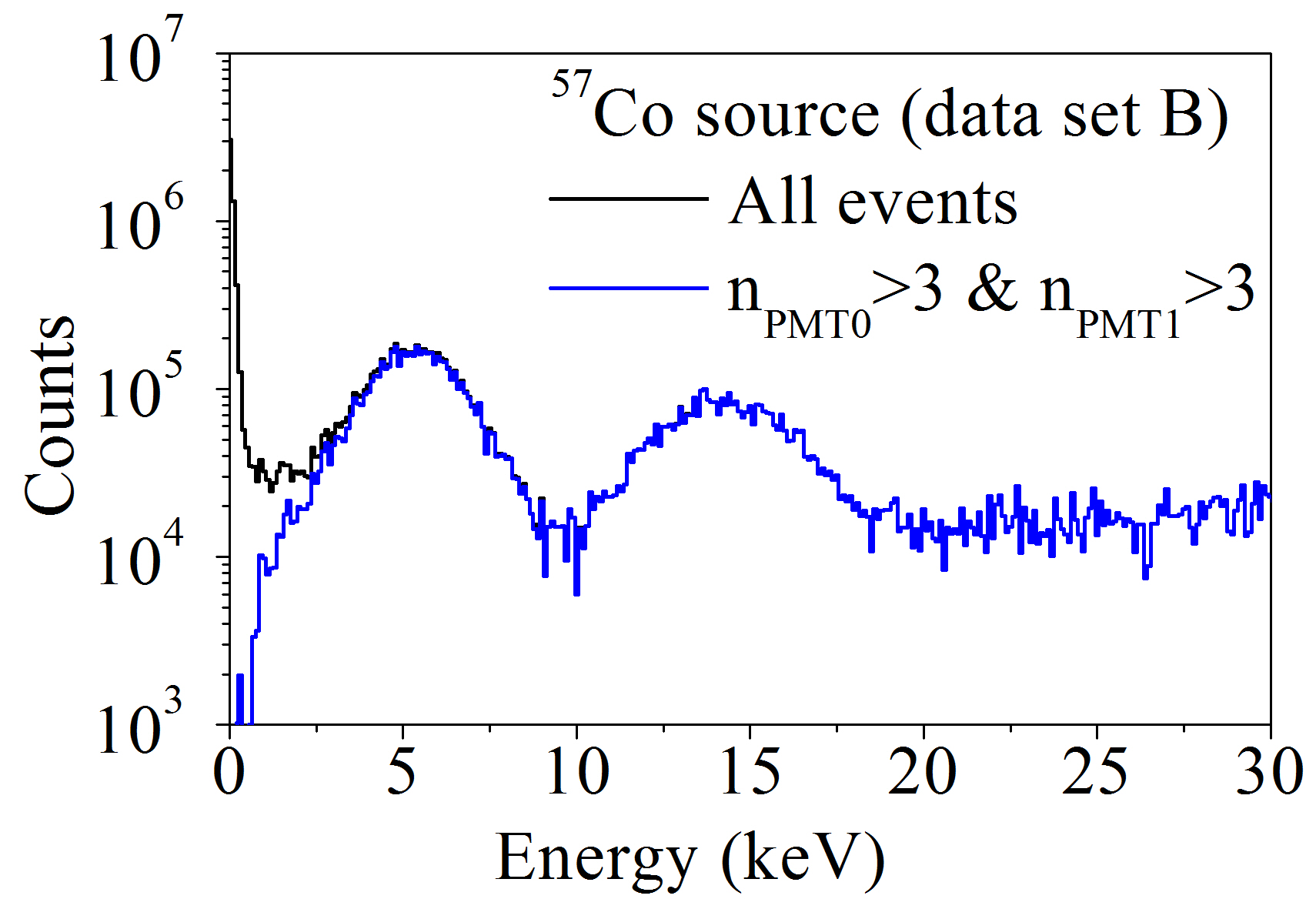}}
\subfigure[]{\includegraphics[width=0.3\textwidth]{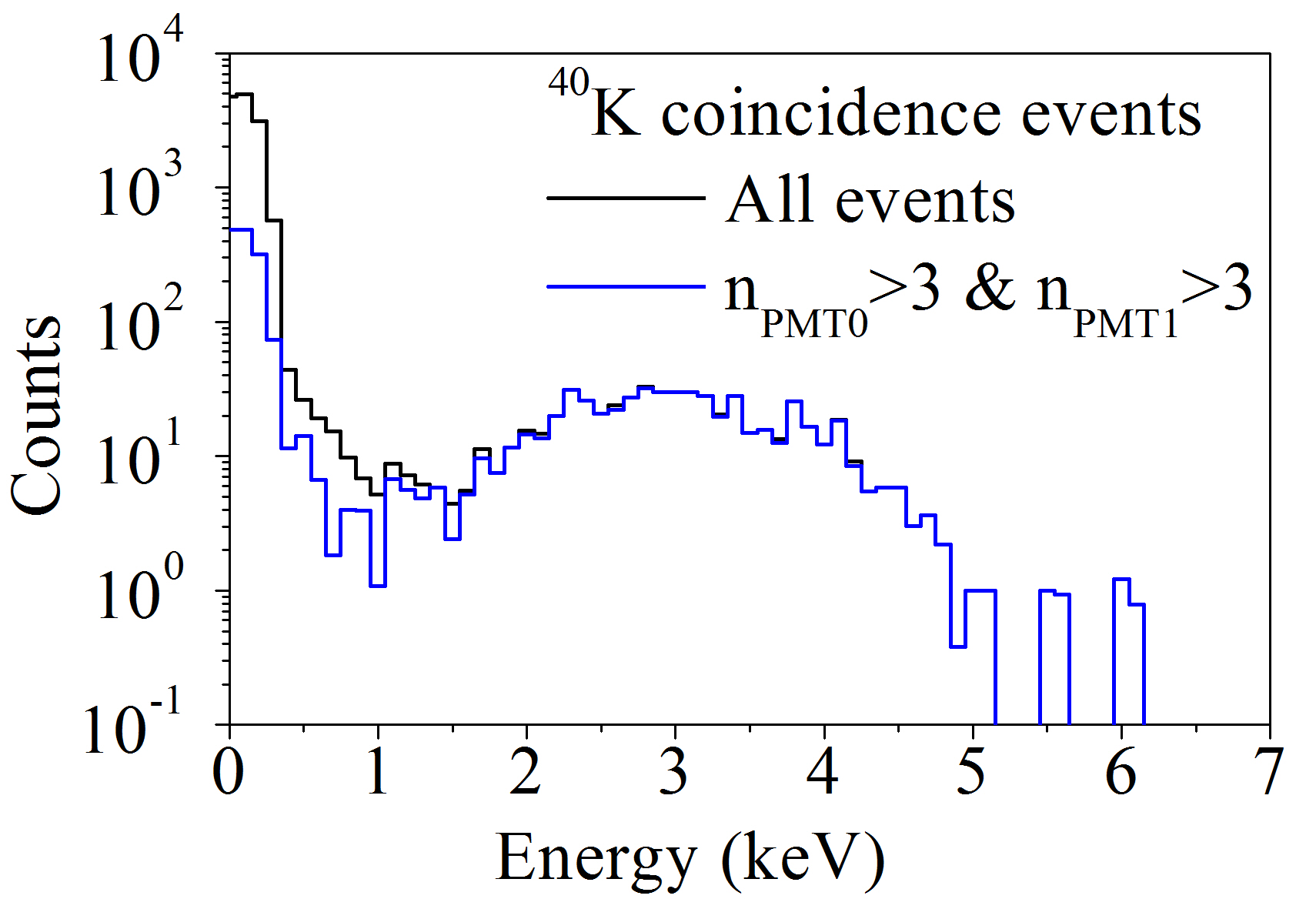}}
\centering \caption{\it Low energy spectra of data set A~(a) and~B~(b) corresponding to a $^{57}$Co calibration and low energy spectrum coincident with the $^{40}$K high energy gamma during a dedicated set-up (c), before (black) and after (blue) the application of filter~3.}
\label{fig:co57n}
\end {figure}

The low energy background spectra before and after the application of filter~3 are shown in Figure~\ref{fig:spectran} for data sets~A and~B. Also, the spectra of the events rejected by this filter are shown. Cerenkov light in the PMT glass produces very fast pulses with energies up to 20\,keVee that are rejected by this filter. The spectra corresponding to the events rejected by this filter for data sets~A and~B, although sharing some features, are different, supporting the hypothesis of their PMT origin. In particular, data set~B PMTs present larger dark current, higher radioactive contamination, and Cerenkov light emission in the PMT glass is expected to be produced while PMTs used in data set A are not made with glass. This could explain the much higher events rate rejected by filter~3 in that data set.

\begin {figure}[ht]
\subfigure[]{\includegraphics[width=0.35\textwidth]{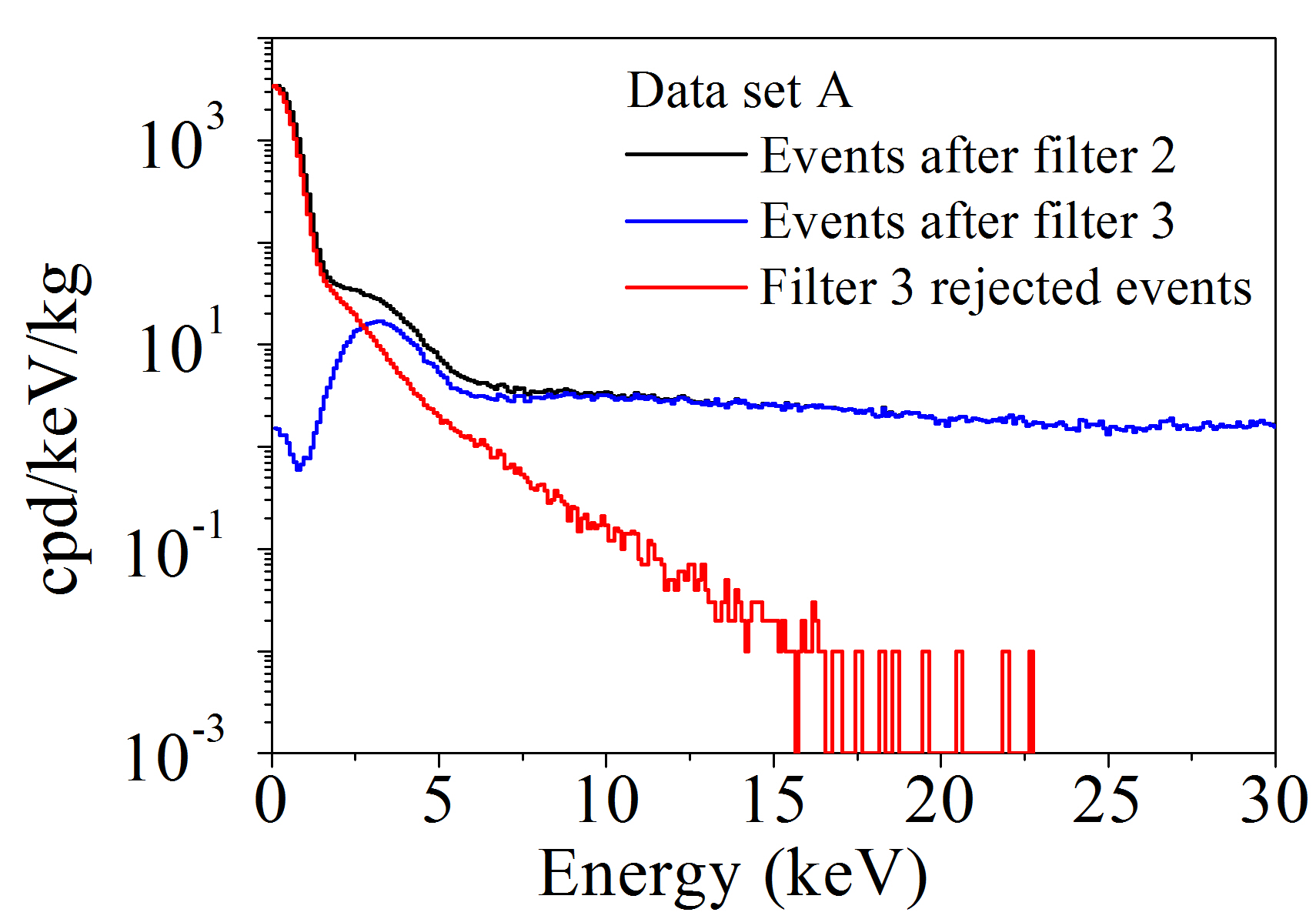}}
\subfigure[]{\includegraphics[width=0.35\textwidth]{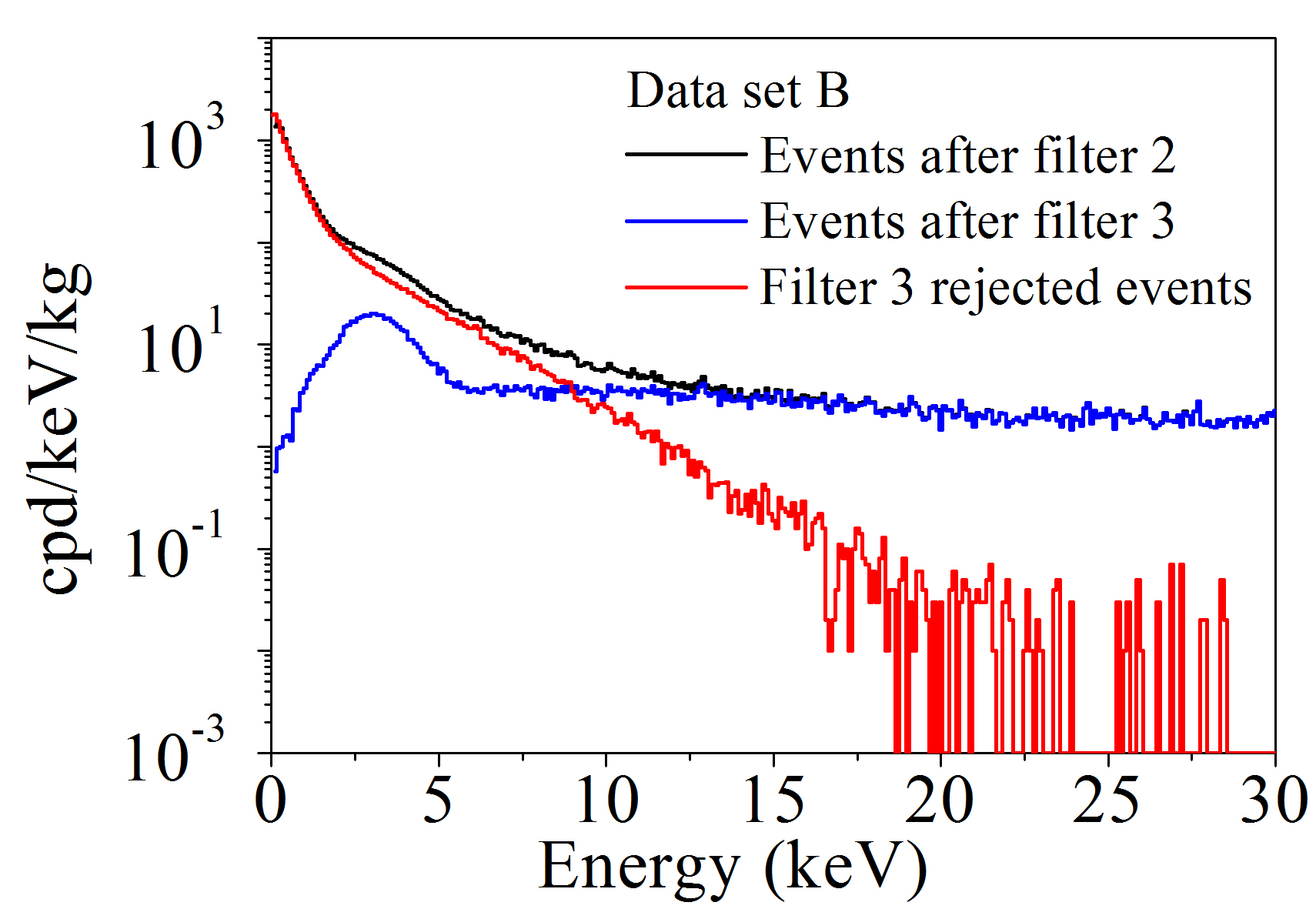}}
\centering \caption{\it Low energy spectra of data set A~(a) and~B~(b) after the application of filter~2 (black), and after filter~3 (blue) without applying any efficiency factor correction. Events rejected by filter 3, because the number of photoelectrons is below 3 in at least one of the two PMT signals, are shown in red.}
\label{fig:spectran}
\end {figure}

	\item \textbf{Events faster than NaI(Tl) bulk scintillation.}
	With the purpose of rejecting events clearly faster than typical NaI(Tl) bulk scintillation pulses, different parameters have been studied. Among them, so called P1s has shown to perform well in the discrimination of anomalous fast events. P1s is defined by the ratio between the addition of the pulse areas from 100 to 600\,ns for the two PMT signals, and the addition of the areas from pulse onset:

\begin{equation}
P1s = \frac{Area1(100-600ns)+Area2(100-600ns)}{Area1(0-600ns)+Area2(0-600ns)}
\end{equation}

This parameter is expected to be around 0.7 for the NaI(Tl) scintillation events even though it shows a slight dependence on the energy (see Figure~\ref{fig:p1vseCo57} where this parameter is shown for $^{57}$Co calibration data up to 100\,keV).

\begin {figure}[ht]
\subfigure[]{\includegraphics[width=0.3\textwidth]{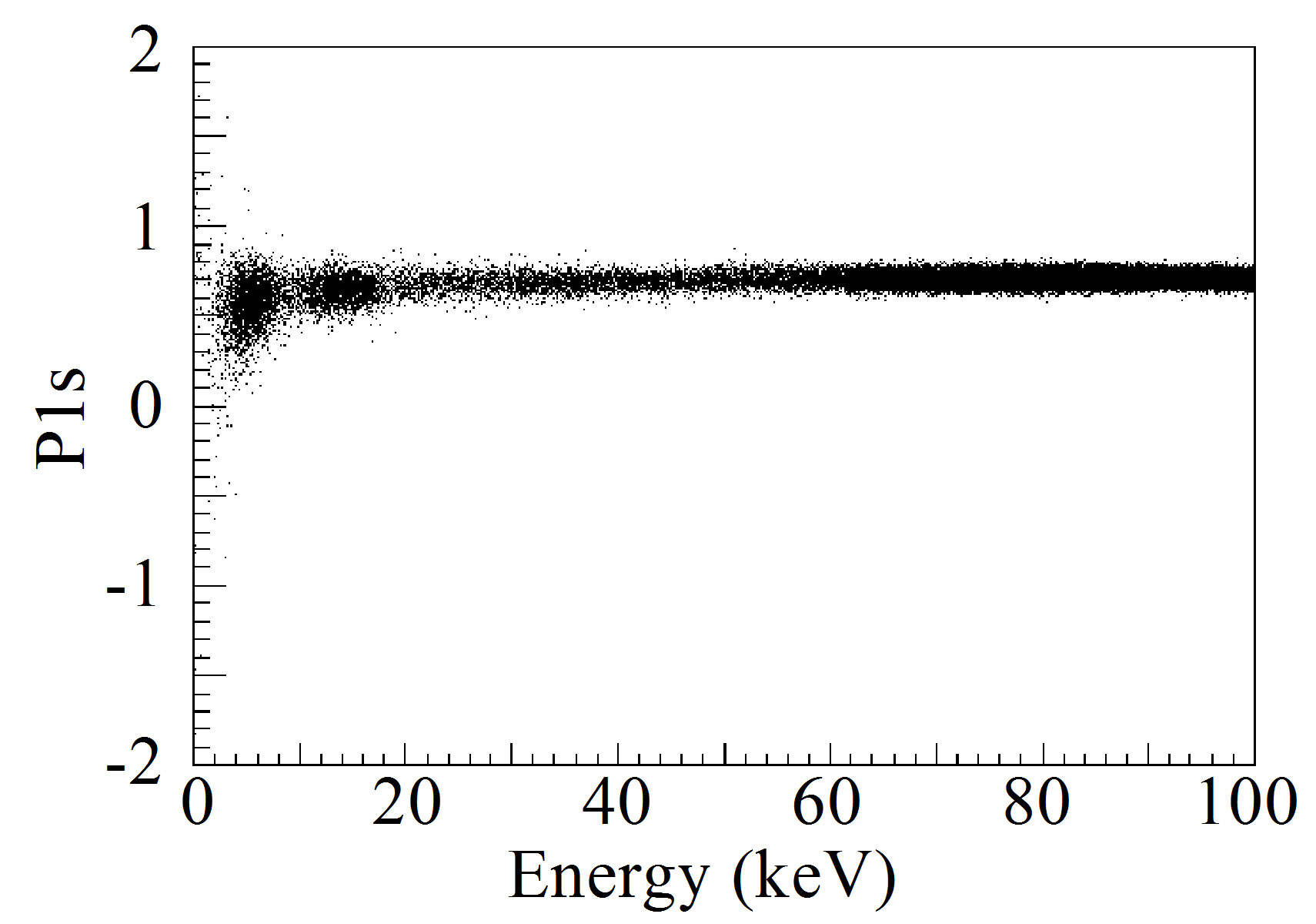}}
\subfigure[]{\includegraphics[width=0.3\textwidth]{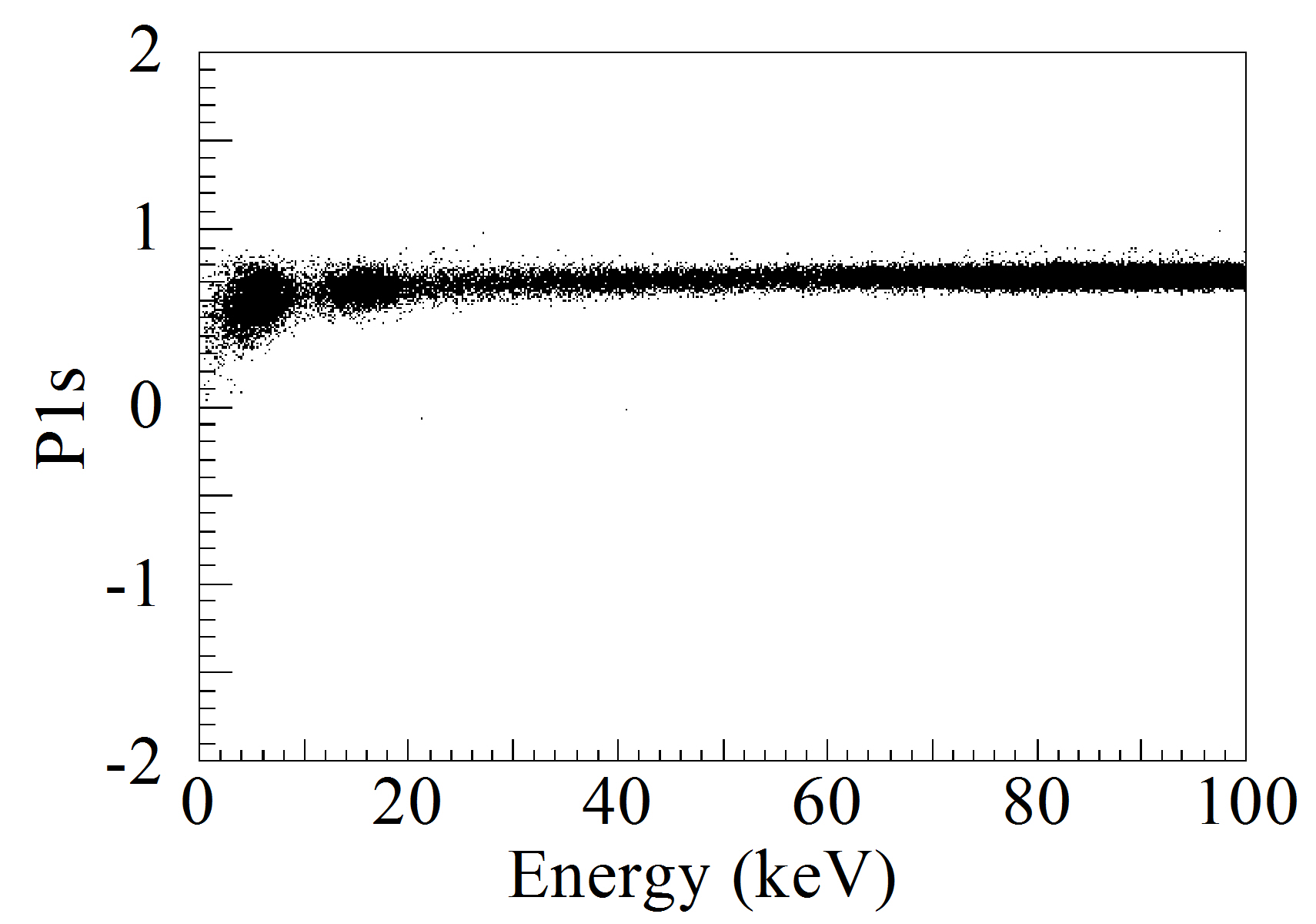}}
\centering \caption{\it P1s parameter distribution as a function of the energy for $^{57}$Co calibration data, after the application of the 3 filters explained in the text. Data in (a) correspond to data set~A and in (b) to data set B.}
\label{fig:p1vseCo57}
\end {figure}

In order to calculate the efficiency of a filter based on this parameter for the acceptance of bulk scintillation events, data from $^{57}$Co and $^{109}$Cd calibrations have been studied for data set A: mean values and standard deviation of P1s in different energy windows (1\,keV width) have been calculated by fitting to a gaussian function the P1s parameter distribution. During data set~B no $^{109}$Cd calibration was available, but data from $^{57}$Co and $^{133}$Ba were used instead. In principle, the values of this parameter should be characteristic of the NaI(Tl) scintillation and do not depend on the special data set features. However, small differences have been observed in the P1s values corresponding to data from data sets~A and~B that could be attributed, for instance, to the better resolution of the MATACQ. An analysis energy threshold of 2\,keVee is imposed hereafter because there are not enough good calibration events below such an energy to allow the definition of an useful acceptance region.

Two filtering criteria have been applied. In the first one, an acceptance region of good scintillation events at 97.7\% is defined by selecting events having P1s value larger than the mean minus 2\,$\sigma$ at every energy window. In the second one, a constant cut of P1s$>$0.4 is selected, and the corresponding efficiency factor calculated in every energy window, and shown in Figure~\ref{fig:eff}. Figure~\ref{fig:p1vse} shows the distribution of the P1s parameter as a function of the energy for background data together with the mean P1s value derived from calibration data and the 2 cut values applied for the filtering, as a function of the energy. A population of events with faster scintillation constant than the typical of NaI(Tl) is clearly observed in both data sets.

\begin {figure}[ht]
\includegraphics[width=0.3\textwidth]{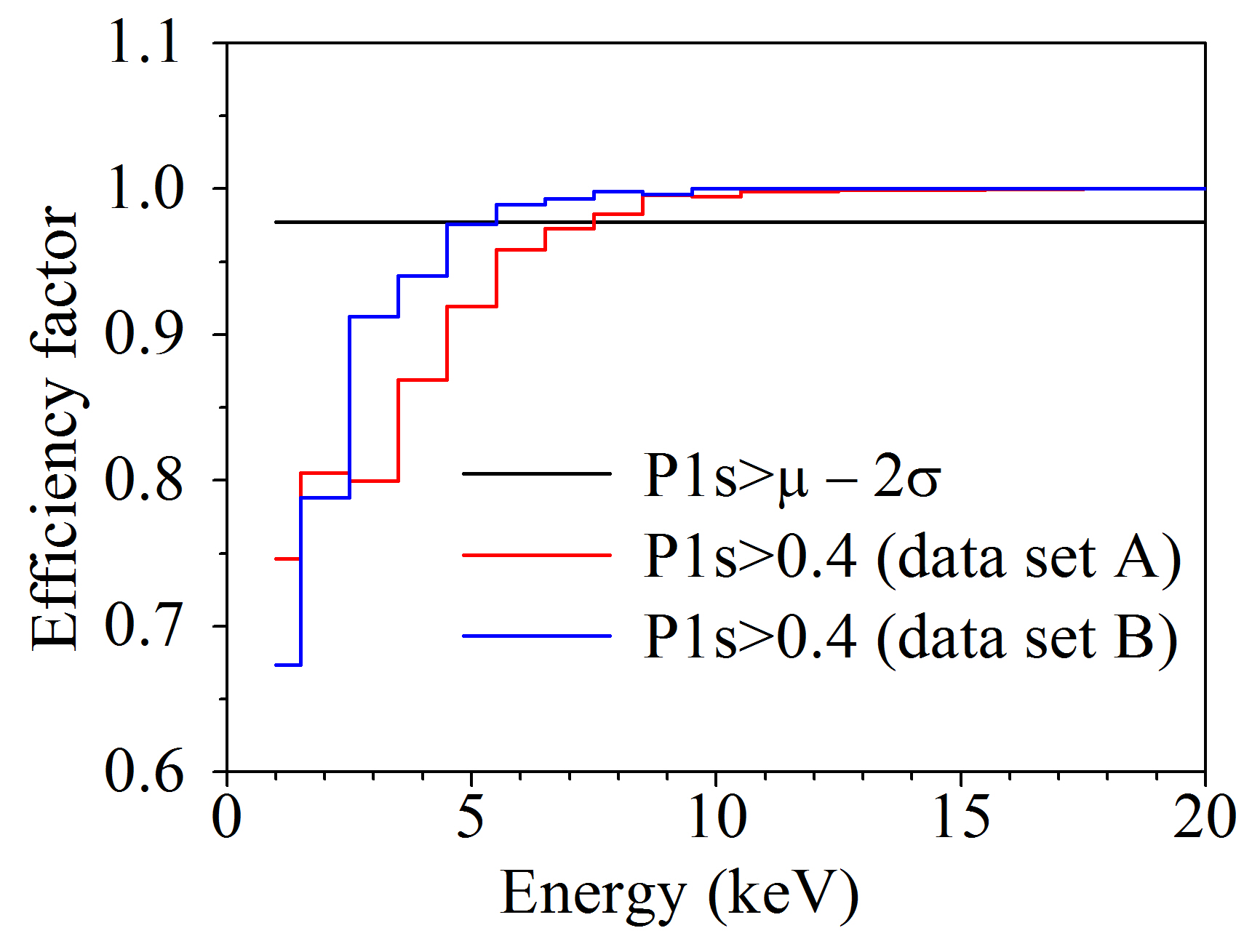}
\centering \caption{\it Efficiency factor derived from calibration data with gamma sources corresponding to filter 4 for the two different definitions of the acceptance region: a constant acceptance factor at 97.7\% (in black) and a cut in a constant value of P1s (in red for data set A and in blue for data set B).}
\label{fig:eff}
\end {figure}

\begin {figure}[ht]
\subfigure[]{\includegraphics[width=0.29\textwidth]{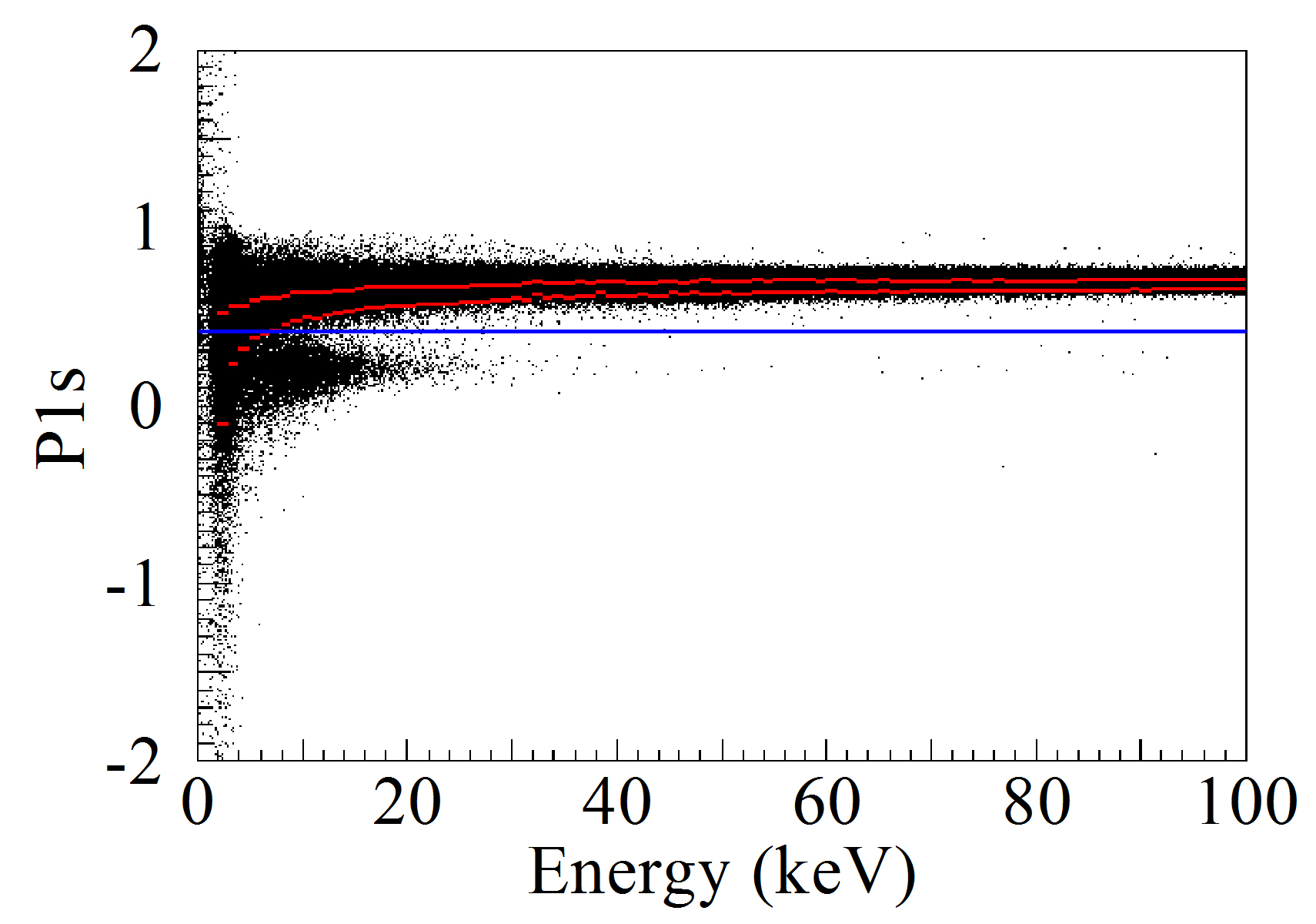}}
\subfigure[]{\includegraphics[width=0.29\textwidth]{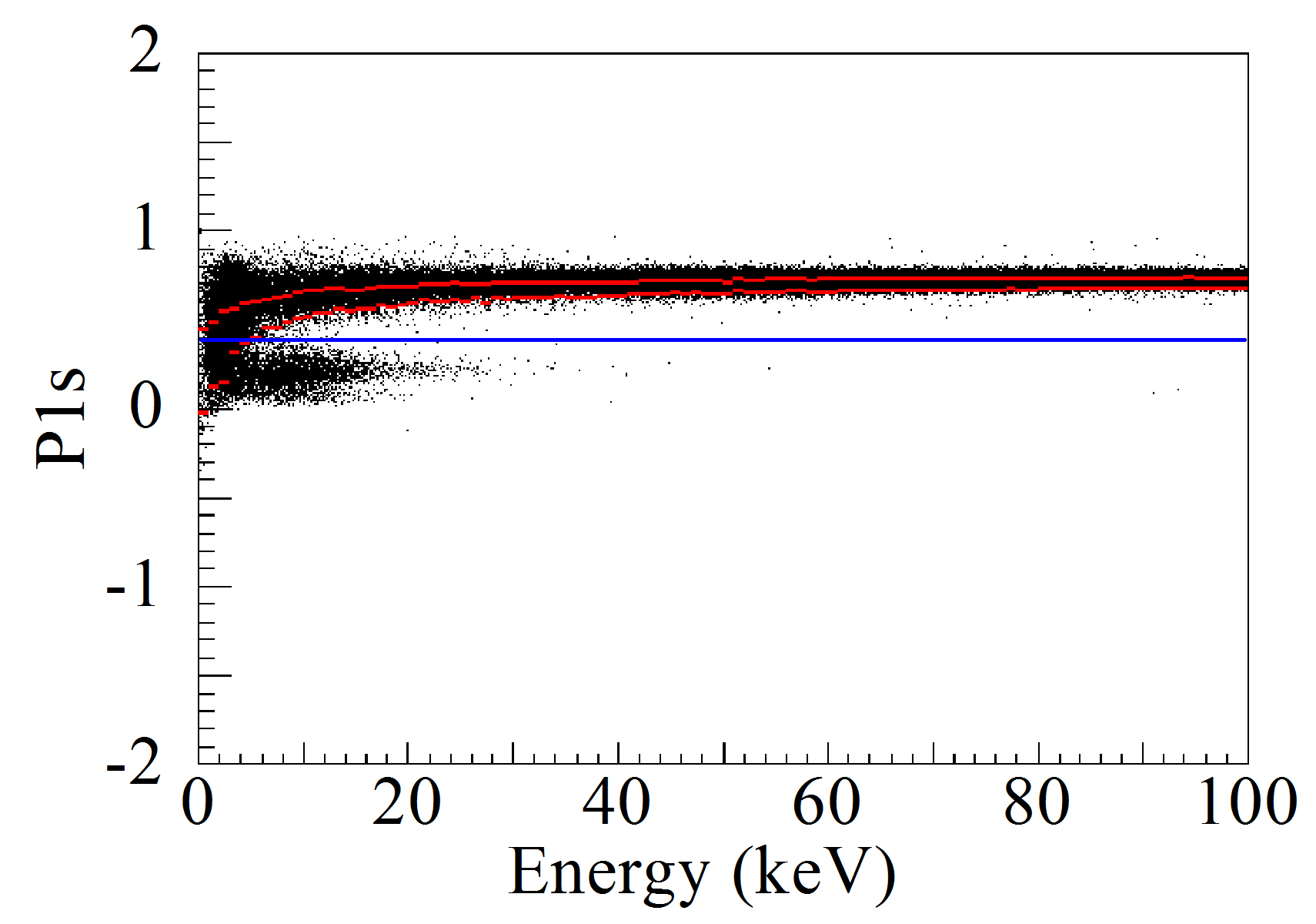}}
\subfigure[]{\includegraphics[width=0.29\textwidth]{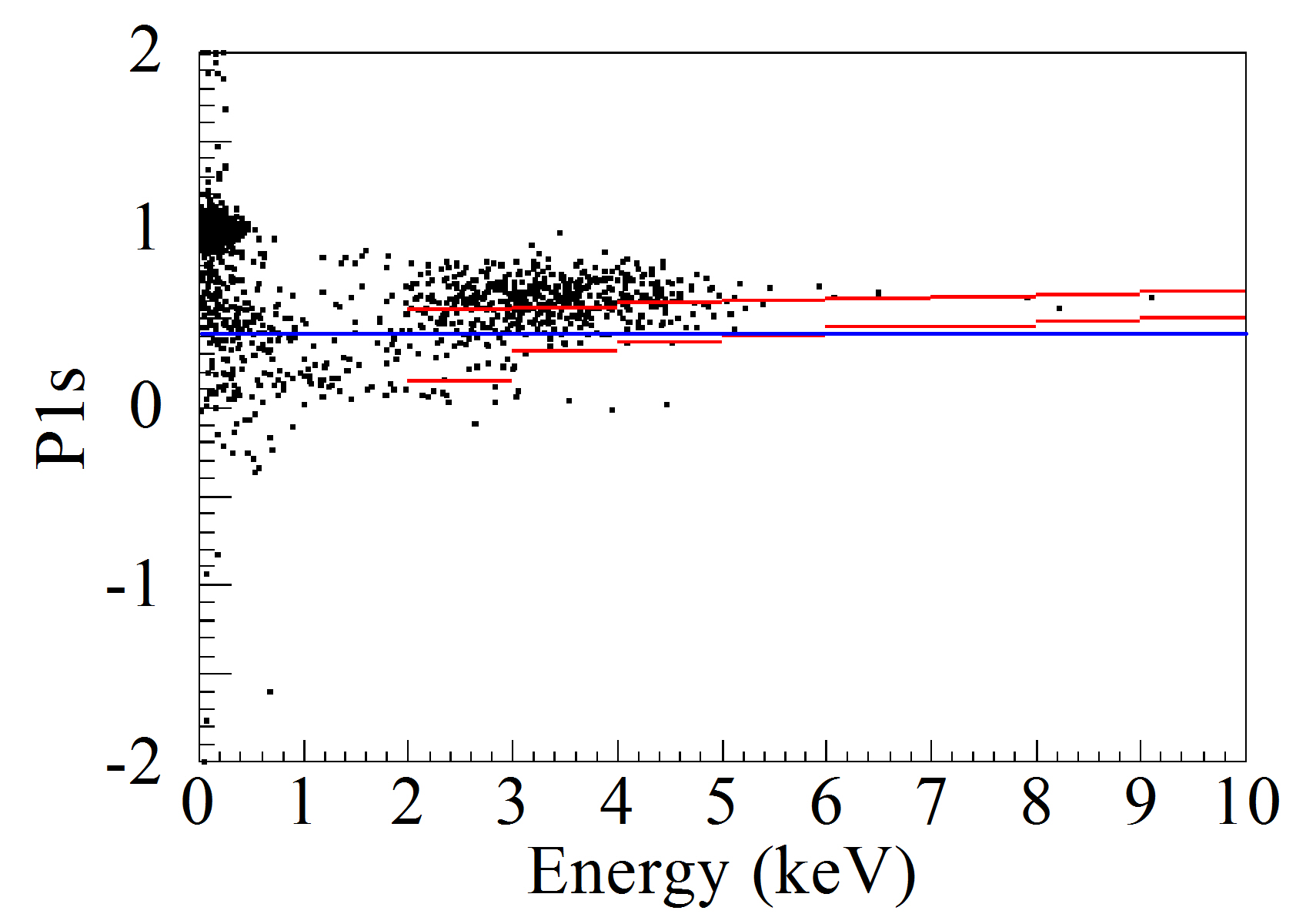}}
\centering \caption{\it P1s parameter distribution versus energy for events passing all the 3 filters referred in the text corresponding to background data from data sets~A~(a) and~B~(b), as well as for 3.2\,keV events coming from the internal $^{40}$K disintegration and selected by the coincidence with the high energy gamma (c). The upper red line represents the P1s mean value obtained from the calibrations and the lower red line the cut value chosen for filter~4 in order to have an acceptance of 97.7\% for bulk scintillation in NaI(Tl) events. The blue line represents the constant cut P1s$>$0.4.}
\label{fig:p1vse}
\end {figure}

The goodness of the so calculated efficiencies for this filter is checked by using the 3.2\,keV events coming from the internal $^{40}$K disintegration, which have been selected by the coincidence method. It can be observed in Figure~\ref{fig:p1vse}.c that these events are not rejected by this filter, as expected for bulk NaI scintillation events. In the energy region from 2 to 10\,keV, there are 553 events. After applying the filter with the 97.7\% acceptance criterion, 537 events remain that correspond to 550 events after applying the efficiency correction. When applying the filter with a constant cut at P1s$>$0.4, 505 events remain that correspond to 576 events after applying the efficiency correction, confirming that both filtering criteria are not losing good scintillation events down to 2\,keVee, and even pointing at an underestimate of the efficiency for the second criterion.

\begin {figure}[ht]
\subfigure[]{\includegraphics[width=0.35\textwidth]{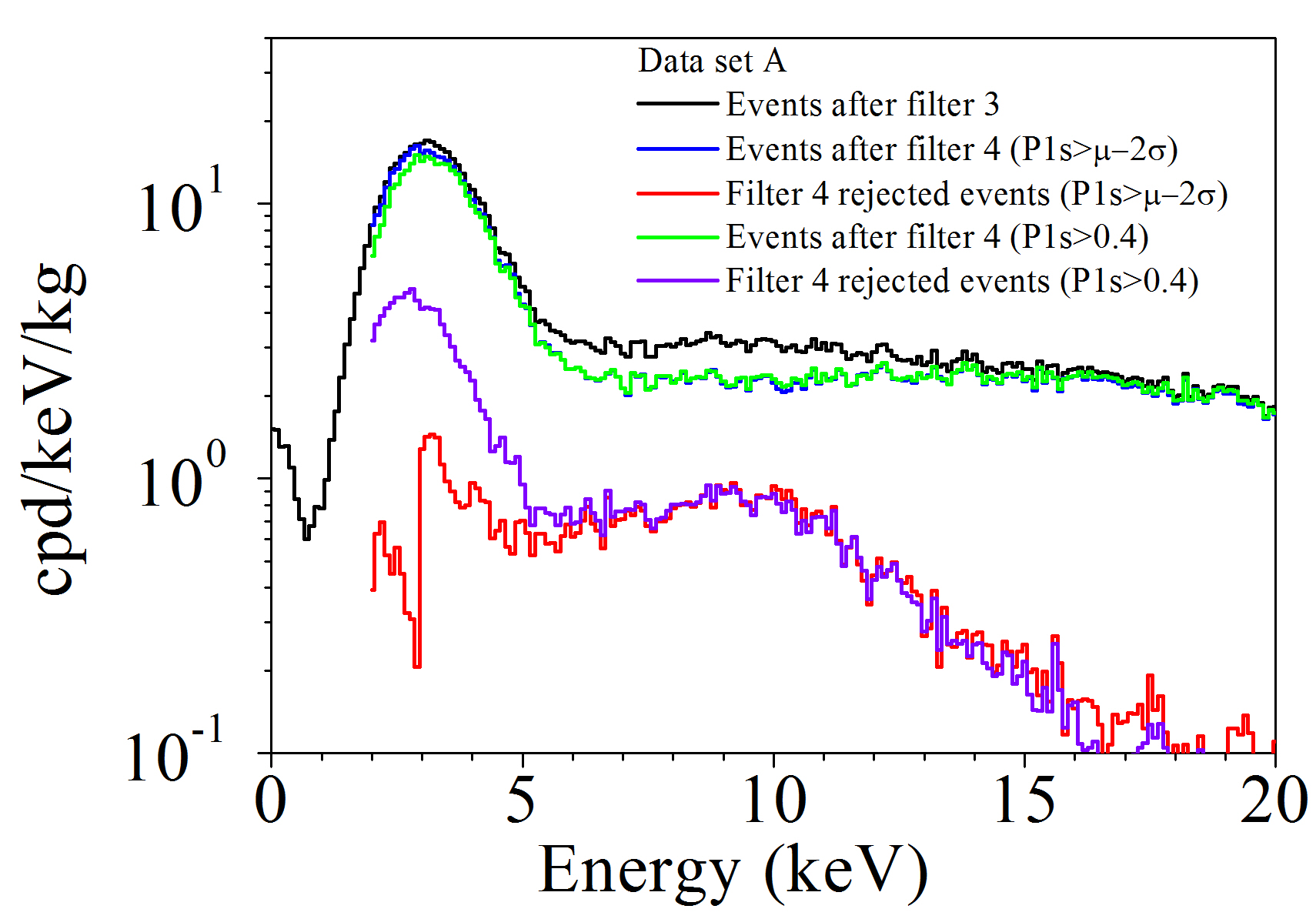}}
\subfigure[]{\includegraphics[width=0.35\textwidth]{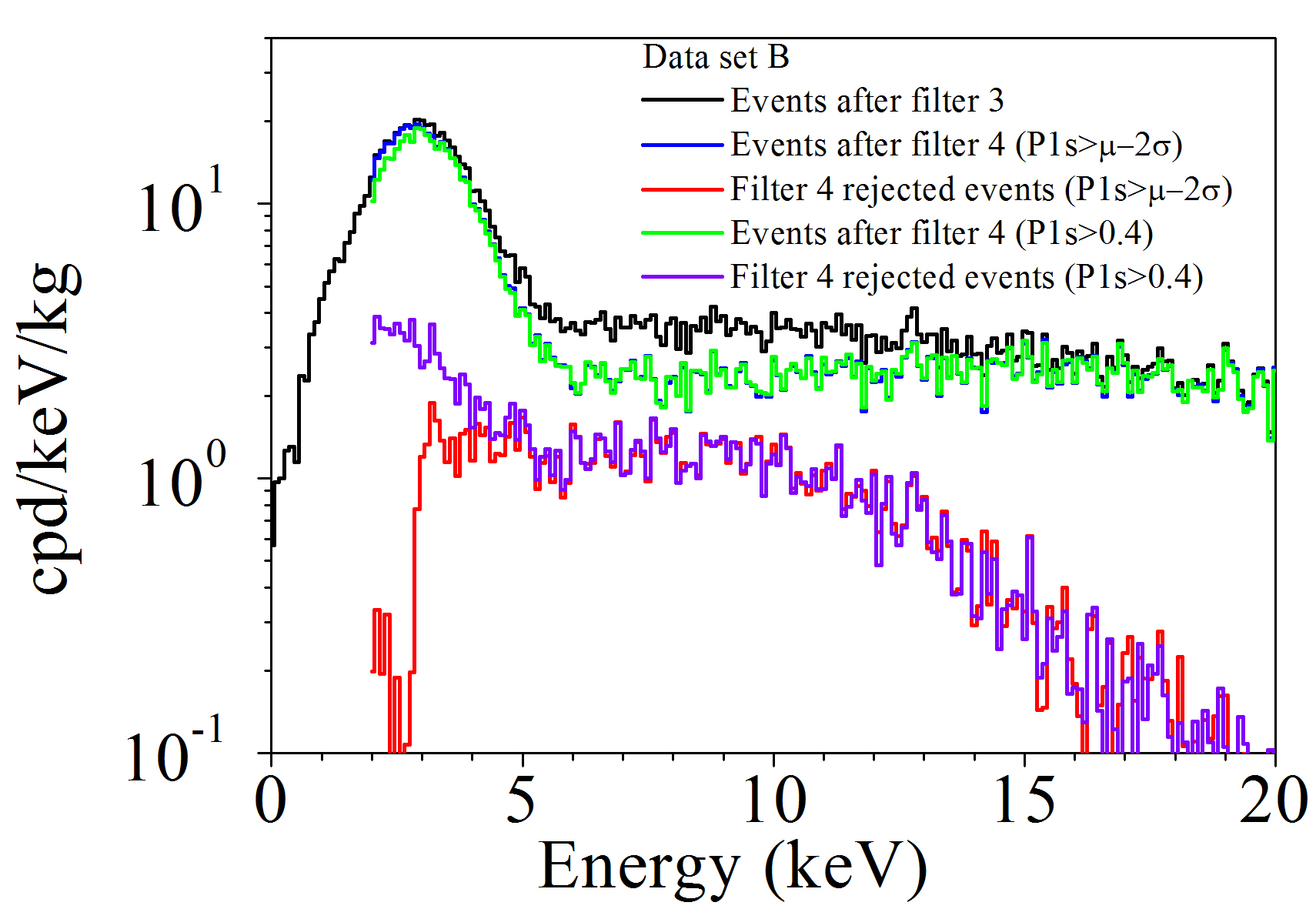}}
\centering \caption{\it Low energy spectra corresponding to data sets A (a) B (b). Events remaining after filter~3 are shown in black, and after applying filter~4 in blue (97.7\% acceptance cut) and green (P1s$>$0.4 cut). Events rejected by filter~4 are also shown in red (2$\sigma$ cut) and violet (P1s$>$0.4 cut).}
\label{fig:badp1}
\end {figure}

Figure~\ref{fig:badp1} shows the spectra obtained after applying this filter with the two different criteria to the background data, and Figure~\ref{fig:zoomp1} a zoom into the low energy region. It can be also observed in Figure~\ref{fig:badp1} the spectra of the events rejected by this filter.

From 2 to 3\,keVee the first criterion is so conservative that it almost does not reject events at all. However, when applying a more effective rejecting criterion, it can be observed that above 3\,keVee both criteria result in the same filtered spectrum. Spectra corrected by the corresponding efficiencies are shown in more detail in Figure~\ref{fig:zoomp1}. The mean values and standard deviation of P1s of the non-bulk NaI(Tl) scintillation low energy events in the background have been calculated by fitting to two gaussian functions the P1s parameter distribution (bulk and non-bulk NaI(Tl) scintillation). From this estimation, we can conclude that rejection of the anomalous events is larger than 99\% above 8\,keV for both criteria, but it differs near the threshold: in the 2-3\,keV region we estimate rejection of 23\% for the 2$\sigma$ cut, whereas of 89\% for the P1s$>$0.4 cut.

\begin {figure}[ht]
\subfigure[]{\includegraphics[width=0.35\textwidth]{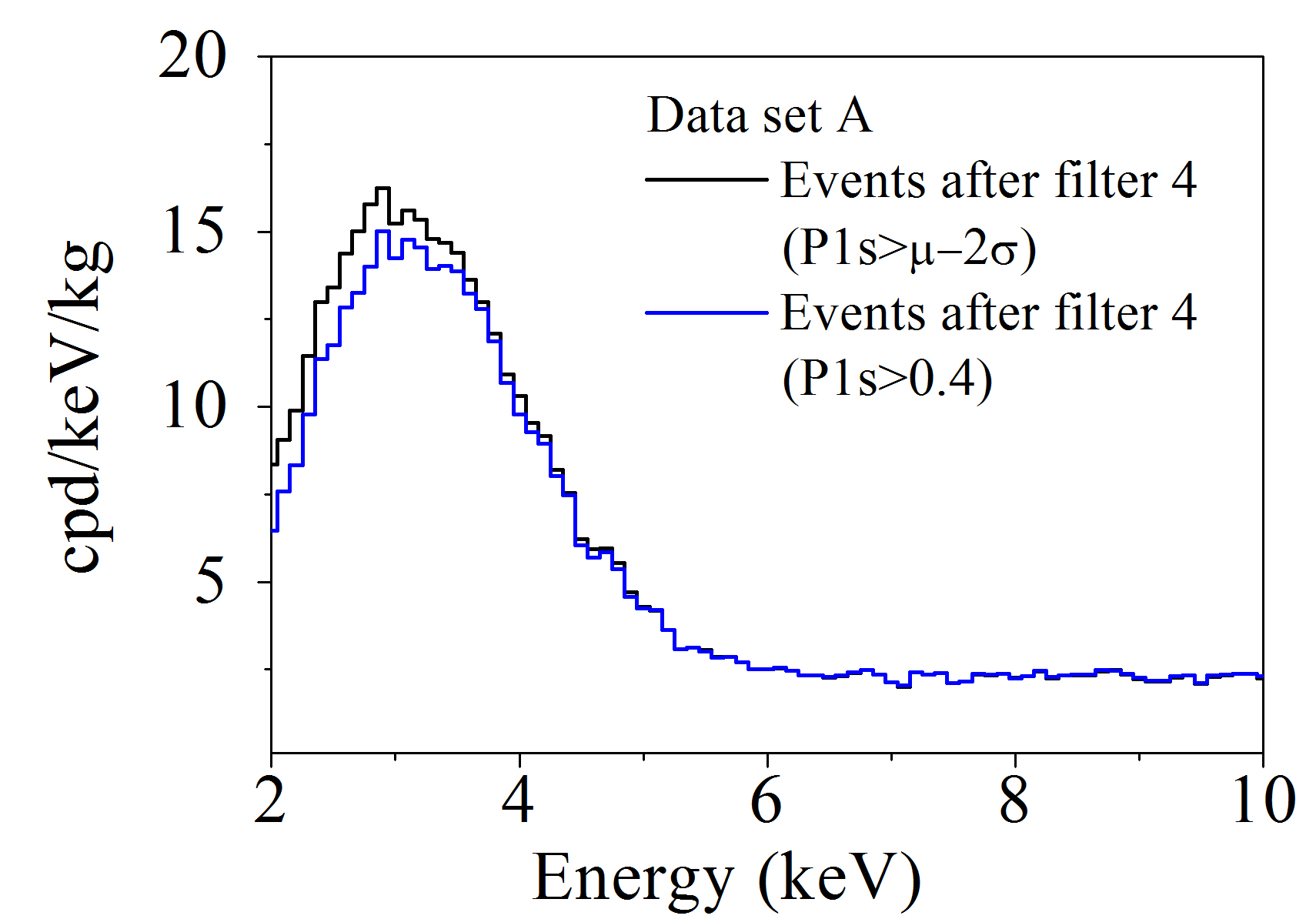}}
\subfigure[]{\includegraphics[width=0.35\textwidth]{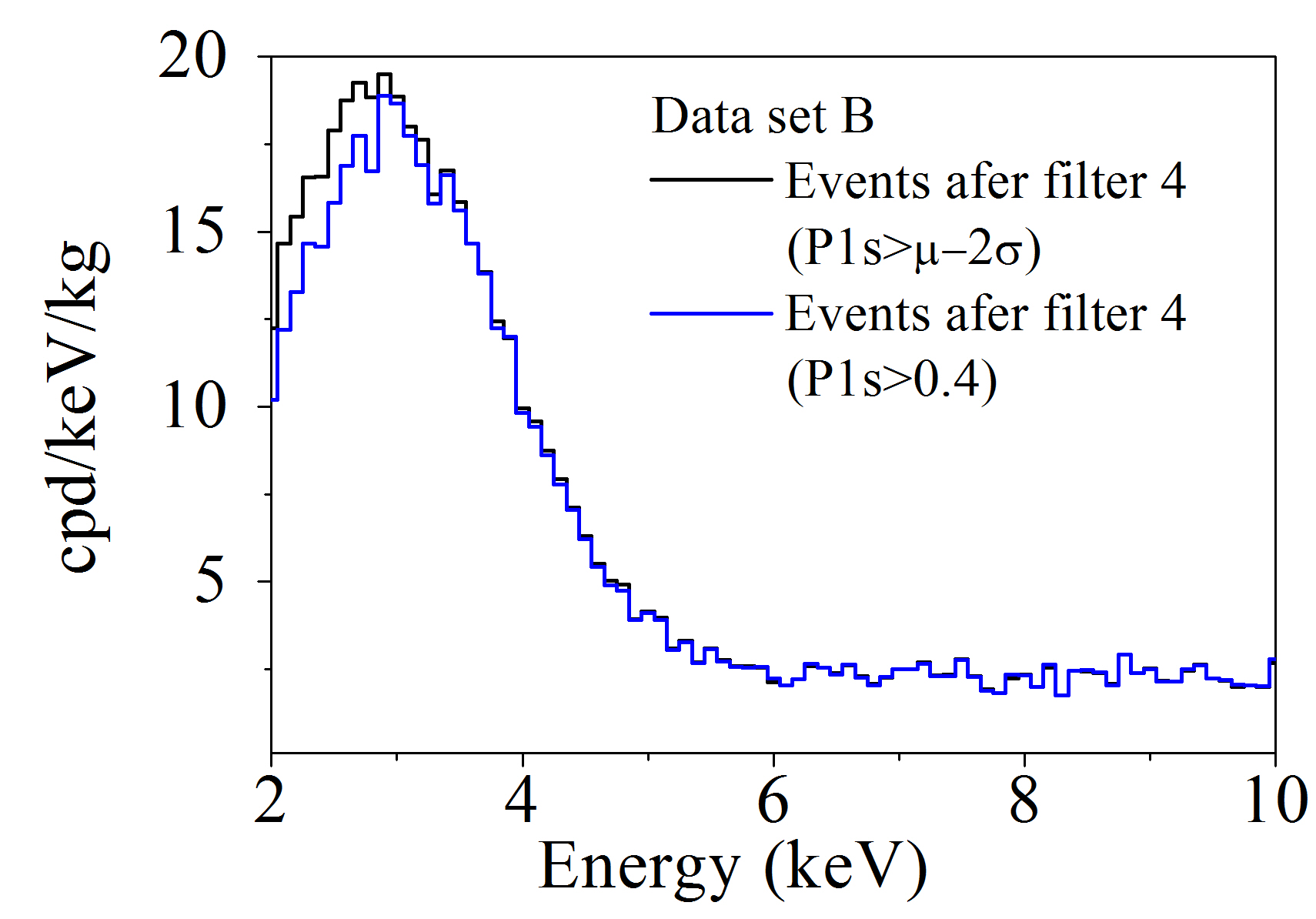}}
\centering \caption{\it Very low energy spectra of events remaining after filter~4 in black (97.7\% acceptance cut) and blue (P1s$>$0.4 cut) corresponding to data sets A (a) B (b) corrected by the corresponding efficiencies.}
\label{fig:zoomp1}
\end {figure}

The origin of these rejected events could be related to surface energy depositions in the NaI(Tl) crystal from isotopes implanted, for instance, after $^{222}$Rn deposition and decay. In fact, in the case of CsI(Tl) crystals it has been recently evidenced that $^{222}$Rn surface deposition originates an anomalously fast event population down to very low energies~\cite{kimsrn} and it has also been proposed as solution to similar fast anomalous events populations identified in old NaI(Tl) dark matter search experiments \cite{gerbier99,cooper,kudryavtsev02}. Another possibility could be related with scintillation in quartz windows, for instance, clearly evidenced in the case of natural quartz~\cite{quartzANAIS}. However, it cannot be completely excluded this effect could also be present with much lower intensity when using synthetic quartz windows.
\end{enumerate}

\begin {figure}[ht]
\includegraphics[width=0.45\textwidth]{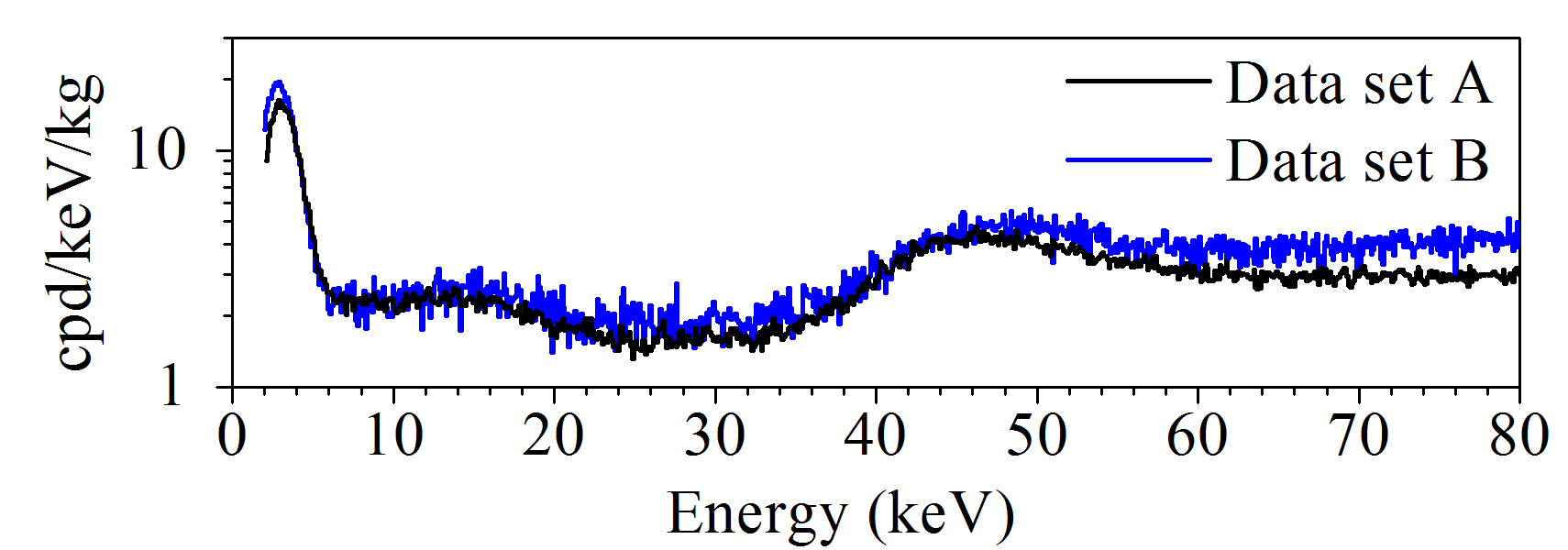}
\centering \caption{\it Low energy spectra of data set~A~(black) and~B~(blue) after having applied all the filters described in text corrected by the corresponding efficiencies. Main differences are observed above 60\,keV and are due to the higher PMTs background contribution in data set B.}
\label{fig:LEcut}
\end {figure}

Figure~\ref{fig:LEcut} shows the spectra of data sets~A and~B after having applied the previously described filters and correcting as explained before live time and efficiency of every selection procedure. For filter~4, the most conservative approach has been selected, trying to minimize rejection, and the 2$\sigma$ cut has been applied. The 3.2\,keV line coming from the $^{40}$K decay dominates the low energy spectrum (from 2 to 6\,keV) in both data sets. From 6\,keV up to 30\,keV a rate of 2-3\,cpd/keV/kg is measured with both PMT models, which is on the limit of the ANAIS experiment requirements. Although the PMTs used in both data sets have very different radiopurity levels~\cite{tesisClara}, differences in the backgrounds observed up to 45\,keV are minimal. This hints at a very low contribution from PMT contamination in the low energy range, as our simulations  suggested~\cite{ANAISbkg}. The difference in the 2-3\,keV region is probably due to non-rejected spurious events having the origin in the PMTs, because of the different PMT models were used in both data sets and the filter chosen. New ultrapure NaI crystals from Alpha Spectra taking data in the LSC (ANAIS-25 set up) will contribute also to conclude that radiopurity levels of R6956MOD PMTs could be enough to reach ANAIS goals in terms of background at very low energy even without using light guides.

\section{Neutron calibration}
\label{fith}
Neutron interactions are relevant for a dark matter experiment because they produce nuclear recoils of the target constituent nuclei as the WIMPs do. Nuclear recoil events in NaI are faster than $\beta / \gamma$ ones~\cite{gerbier99,miramonti02,bernabei96,kudry99}, however at low energies there is only a slight difference in pulse shape and particle discrimination cannot be done on an event by event basis.

\begin {figure}[ht]
\includegraphics[width=0.4\textwidth]{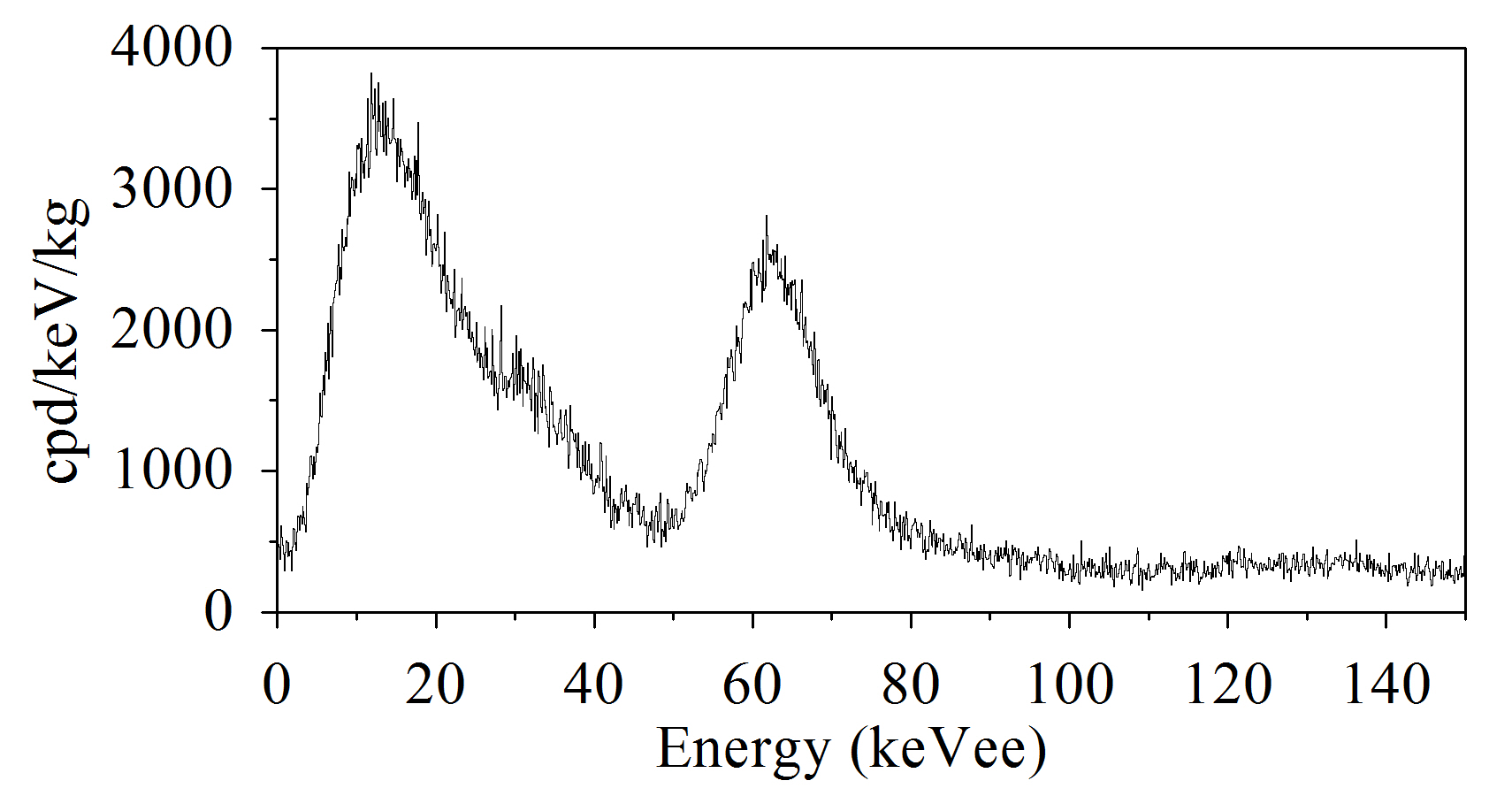}
\centering \caption{\it $^{252}$Cf neutron source calibration spectrum taken with PIII. Only low energy (below 150\,keVee) is shown.}
\label{fig:neutrons}
\end {figure}

Unfortunately, neutron sources are not easily brought to an underground facility and ANAIS experiment had only the chance to do a neutron calibration in 2007 using a $^{252}$Cf source. The neutron emission rate of the $^{252}$Cf source was 4$\times$10$^{4}$\,n/s. At that moment, only the PIII was taking data. Because of that, we have reanalyzed those data in order to obtain information about nuclear recoils pulse shape and then, the effect of the previously explained filters on such a population. Figure~\ref{fig:neutrons} shows the low energy spectrum obtained with the $^{252}$Cf source. Threshold effects are observed at energies lower than 10\,keVee. According to specific Geant4 simulations, up to 30\,keVee most of the events in the spectrum can be attributed to Na nuclear recoils.

\begin {figure}[ht]
\subfigure[]{\includegraphics[width=0.3\textwidth]{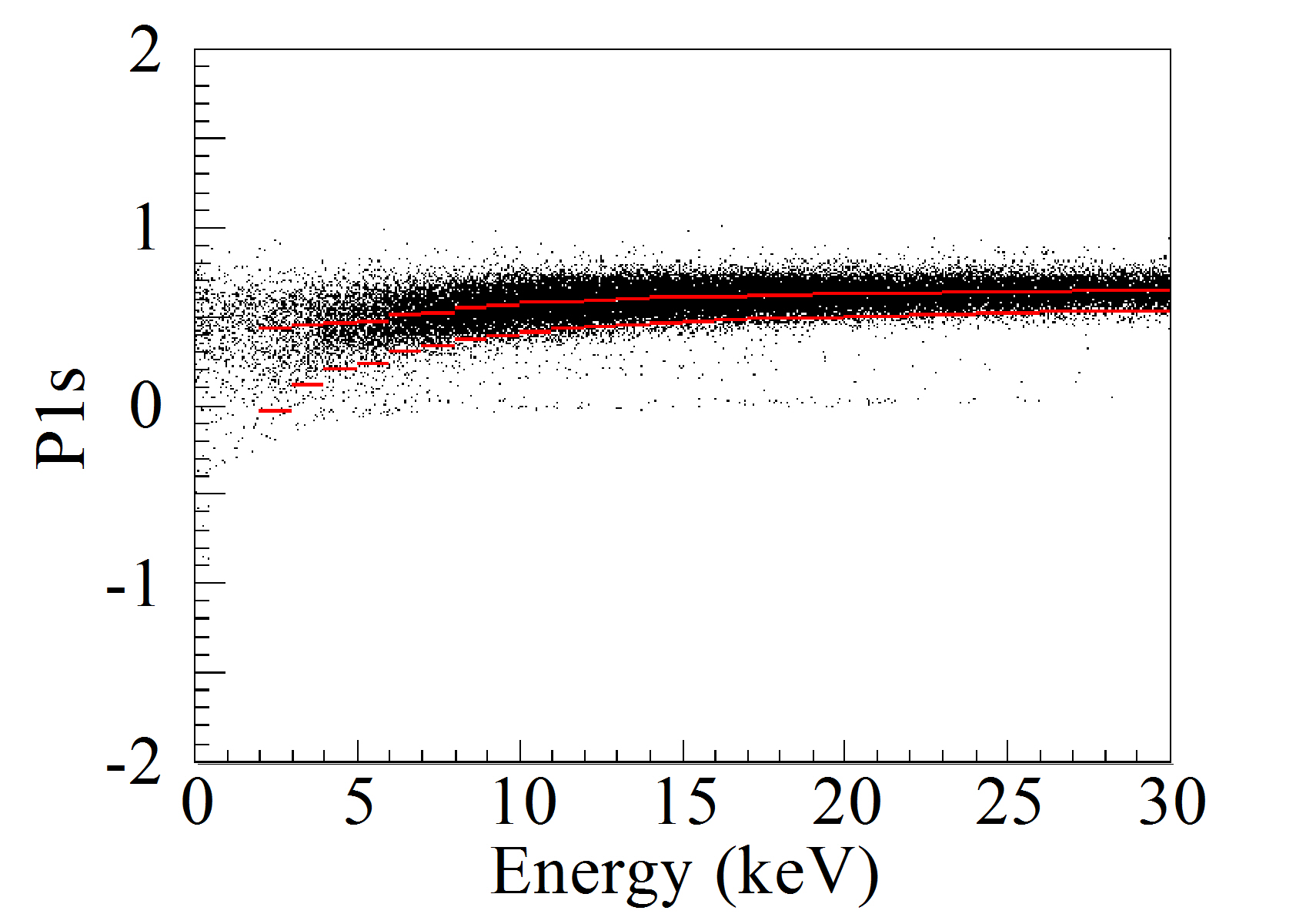}}
\subfigure[]{\includegraphics[width=0.3\textwidth]{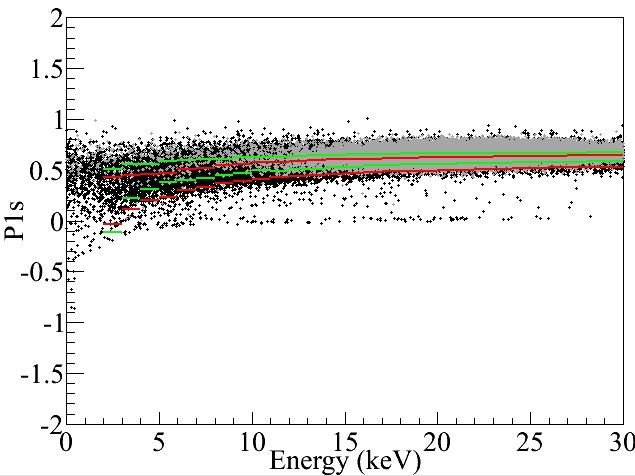}}
\centering \caption{\it (a) P1s parameter versus energy for $^{252}$Cf calibration data taken with PIII. The upper red line represents the P1s mean value obtained in 1\,keV width windows and the lower red line the cut value chosen to define the acceptance region (the mean minus 2\,$\sigma$). (b) $^{252}$Cf calibration (black), $^{109}$Cd\,+\,$^{57}$Co calibrations (gray), P1s mean value and the cut value chosen to define the acceptance region (the mean minus 2\,$\sigma$) for the $^{252}$Cf (red) and $^{109}$Cd\,+\,$^{57}$Co (green) calibrations.}
\label{fig:p1vseneutrons}
\end {figure}

Figure~\ref{fig:p1vseneutrons}.a shows the P1s parameter as a function of the energy for the $^{252}$Cf calibration data, whereas Figure~\ref{fig:p1vseneutrons}.b shows the distribution of the same parameter for a combination of data from $^{109}$Cd\,+\,$^{57}$Co (in blue) and $^{252}$Cf (in black). The mean value of the P1s parameter has been analyzed in 1\,keV width windows, as it was done in section 4 for the $\beta / \gamma$  events from $^{109}$Cd and $^{57}$Co calibrations. No filtering has been applied to the $^{252}$Cf calibration data and some pulse tails and single photoelectron events are present with P1s $\approx$ 0, but they only amount a 0.2\% of the total number of events, and they do not affect to the conclusions derived below. For $^{109}$Cd and $^{57}$Co calibration data, filters 1 to 3 have been applied. Corresponding cut values defining 97.7\% bulk scintillation acceptance regions are shown in red and green, respectively. It can be seen that neutrons are faster than $\beta / \gamma$ particles, but a cut cannot be established to distinguish them event by event.

Figure~\ref{fig:p1vseneutrons} shows clearly that nuclear recoil events would be rejected by the cut on the P1s parameter with an efficiency different than that estimated from beta/gamma calibrations. The cut value chosen with beta/gamma events (lower limit at $2\,\sigma$ confidence level) is equivalent to a lower limit at ($1.59\pm0.37)\,\sigma$ for neutron events, if data from 5 to 30\,keV are averaged. Hence, in the 5 to 30\,keVee energy window we should replace our acceptance efficiency for the selection of events using filter~4 (see section~\ref{forth}) by the values shown in Table~\ref{tab:neutrons} instead of 0.977 in order to conveniently correct for the fraction of nuclear recoils in the acceptance window.

\begin{table}[ht]
\caption{\it Efficiency correction factor (Eff.) for the selection of events using filter~4 that should be considered in order to conveniently correct for the fraction of nuclear recoils in the acceptance window.}
\label{tab:neutrons}
\begin{tabular}{ll}
\hline\noalign{\smallskip}
Energy region & Eff. \\
\noalign{\smallskip}\hline\noalign{\smallskip}
5 - 10\,keV & 0.862\\
10 - 15\,keV & 0.942\\
15 - 20\,keV & 0.949\\
20 - 25\,keV & 0.959\\
\noalign{\smallskip}\hline
\end{tabular}
\end{table}

\appendix

\section*{Summary}

Low energy events are of utmost interest for the ANAIS experiment because dark matter particles are expected to produce very small energy depositions in the detector. For this reason, a very good knowledge of the detector response function for real scintillation events in the active volume of the detector and a good understanding and characterization of other anomalous or noise event populations in that energy region are required. Efficiently filtering all the low energy events populations non attributable to dark matter interactions is one of the main issues for ANAIS. Among them, events having scintillation time constants other than NaI(Tl) one or events in coincidence with a signal in the plastic scintillator vetoes should be rejected. Specific protocols to reject such events have been developed and applied to data from \mbox{ANAIS-0} module. We have demonstrated the ability to reject anomalous or spurious event populations and to estimate the corresponding efficiencies or live time corrections down to 2\,keVee. We have shown how to adapt our filtering procedures to a nuclear recoil population as the expected for galactic halo WIMPs interacting in our detector. We have followed a very conservative approach, not maximizing rejection but acceptance. A 2\,keVee energy threshold can be confirmed after applying such protocols to \mbox{ANAIS-0} data, and further improvement is expected in next prototypes, by improving light collection efficiency and increasing the number of reference events in every energy bin. The background below 6\,keVee is clearly dominated by the 3.2\,keV line from $^{40}$K and in the region from 6 to 30\,keVee a rate of 2-3\,cpd/keV/kg has been obtained without light guides for the two PMT models considered. The filtering described in this article can be applied with only slight modifications to ANAIS experiment.

\section*{Acknowledgements}

This work has been supported by the Spanish Ministerio de Econom\'{\i}a y Competitividad and the European Regional Development Fund (MINECO-FEDER) (FPA2011-23749), the Consolider-Ingenio 2010 Programme under grants MULTIDARK CSD2009-00064 and CPAN CSD2007-00042, and the Gobierno de Arag\'{o}n (Group in Nuclear and Astroparticle Physics, ARAID Foundation and C. Cuesta predoctoral grant). C. Ginestra and P. Villar have been supported by the MINECO Subprograma de Formaci\'{o}n de Personal Investigador. We also acknowledge LSC and GIFNA staff for their support.

\section*{References}


\bibliographystyle{elsarticle-num}
\bibliography{Bulk_arxiv2_v2}







\end{document}